\documentclass[a4paper,12pt]{article}
\usepackage{amsmath,amsthm,amsfonts,amssymb,bm,mathrsfs}
\usepackage{enumerate}
\usepackage{graphicx}
\usepackage{natbib}
\usepackage{xcolor}
\IfFileExists{sgame.sty}{\usepackage{sgame}}{}
\usepackage{comment}
\DeclareUnicodeCharacter{2011}{-}

\usepackage{setspace}
\allowdisplaybreaks

\newtheorem{defn}{Definition}
\newtheorem{thm}{Theorem}
\newtheorem{lma}{Lemma}

\newtheorem{rmk}{Remark}
\newtheorem{exam}{Example}
\newtheorem{cro}{Corollary}
\newtheorem{fact}{Fact}

\newtheorem{claim}{Claim}

\newcommand{\cG}{\mathcal{G}}

\newcommand{\cI}{\mathcal{I}}

\newcommand{\ls}{\mbox{Ls}}

\DeclareMathOperator*{\argmax}{argmax}

\usepackage{hyperref}
\hypersetup{hypertexnames=false,colorlinks=true,
            citecolor=blue,
            linkcolor=blue,
            urlcolor=blue,
            bookmarksopen=true,
            pdfstartview={XYZ null null 1.00},
            pdfpagelayout={SinglePage}
}

\newcommand{\revision}[1]{%
  \begingroup%
  #1\endgroup}

\begin{document}
\title{Large Economies with Social Types\footnote{Early versions of this paper have been presented at the 17th SAET Conference, Faro, Portugal, June 2017; the 2017 Asian Meeting of the Econometric Society, Hong Kong, China, June 2017; the 4th PKU-NUS Annual International Conference on Quantitative Finance and Economics, Shenzhen, China, May 2019; the 2019 Nanjing International Conference on Theory of Games and Economic Behavior, Nanjing, China, September 2019;  NSFC Symposium on Young Scientists Fund Projects in Economic Sciences, Kunming, China, September 2023; the 7th World Congress of the Game Theory Society, Beijing, China, August 2024. We thank the participants at these conferences for their valuable comments. Wei He acknowledges financial support from the Research Grants Council of Hong Kong (14506325) and the National Natural Science Foundation of China (No.72122023). Bin Wu acknowledges support from the National Natural Science Foundation of China (No.72203154 and No.72450003). An earlier version of some of the results appeared in Chapter~6 of Bin Wu's Ph.D. thesis \citep{Wu2019}. } }
\author{Wei~He\thanks{Department of Economics, The Chinese University of Hong Kong, Shatin, N.T., Hong Kong. E-mail: hewei@cuhk.edu.hk.}
\and
Yeneng~Sun\thanks{Departments of Economics and Mathematics, National University of Singapore, Singapore 119076. E-mail: ynsun@nus.edu.sg.}
\and
 Bin~Wu\thanks{International School of Economics and Management, Capital University of Economics and Business, Fengtai District, Beijing, 100070, China. E-mail: bwu.econ@gmail.com.}
 \and
 Nicholas~C. Yannelis\thanks{Department of Economics, The University of Iowa, Iowa City, IA 52242-1994, United States.  E-mail: nicholasyannelis@gmail.com.}
}
\date{This version: \today}


\maketitle

\abstract{

This paper develops a new general equilibrium framework for large economies with social externalities, where agents have names as well as social types. Social types define the reference groups (neighborhoods, peers, or status classes) through which externalities operate, and are primitives of the market structure. \revision{We establish that every admissible economy admits a competitive equilibrium if and only if social anonymity holds. This condition requires that observable social types do not fully identify individuals within any non-negligible subeconomy.} The framework extends to generalized large economies with incomplete preferences, indivisible goods, and production,  and thus accommodates nonconvexity in preferences and consumption sets. Our proof replaces the standard aggregate convexification argument with conditional convexification. Applications include peer effects, social comparison, and risk sharing with local social benchmarks.



\smallskip

}

\newpage


\section{Introduction}\label{sec-intro}

Perfect competition is one of the central ideas in economic theory. It not only assumes that agents take prices as given, but also embodies a deeper vision of how markets aggregate decentralized decisions coherently. Since \cite{au64, au66}, the continuum model of large economies has become a canonical formulation of perfect competition: individual choices are decentralized, prices coordinate scarcity, and the market allocation emerges from the aggregation of a society of negligible agents.

Yet agents do not only consume goods; they also live in communities, attend schools, observe peers, compare status, and respond to the behavior of others. For instance, a student's educational investment may depend on the effort of classmates; an agent's housing or durable-goods choice may be shaped by the average standard in her neighborhood; residents may evaluate their own consumption against that of an aspiration group. These externalities are common features of markets in a society, which have been well documented in the literature.\footnote{There is a substantial literature studying economic settings with social interactions and peer effects. Intuitively, an individual's payoff from an activity may depend on the behavior of her peers; see, for example, \cite{Manski1993}, \cite{BD2001} and \cite{BBDI2011}. In many settings, agents may exhibit ``keeping up with the Joneses'' preferences, in which they react to the social pressure from their reference group; see, for example, \cite{Duesenberry1949}, \cite{Abel1990}, \cite{LjungqvistUhlig2000}, and \cite{HopkinsKornienko2004}. }


In a large society, it is implausible that an agent observes the complete allocation or compares herself with every individual. Social comparisons are instead commonly mediated by benchmarks attached to socially meaningful categories. This practice is reflected in public statistics that report various outcomes at regional, occupational, local, or other group levels.\footnote{Examples include regional GDP per capita, the OECD Regional Well-Being indicators, the UNDP Human Development Index, occupation-by-area wage estimates such as the U.S. Bureau of Labor Statistics' Occupational Employment and Wage Statistics, and the FHFA House Price Index. These measures illustrate the widespread practice of summarizing complex individual outcomes at socially meaningful levels.} These summary indicators can serve as aggregate benchmarks conditional on a social category. We refer to the information defining such a category as an agent's social type. It may represent a location, community, school, occupation, cohort, wealth group, status class, network position, or a combination of several attributes. Once such social forces are introduced, the organization of externalities is part of the economic environment. A model must specify not only whose choices matter, but also at what level those choices are observed, aggregated, and made payoff relevant. The social-type structure should therefore be treated as a primitive of the market, on par with commodities, endowments, and preferences.


This observation raises a natural question: how should large economies with externalities be modeled? The question is particularly delicate because externalities create new difficulties for equilibrium existence. Prices still coordinate scarcity and agents remain individually negligible, but the allocation also creates the social environment in which agents evaluate their own choices. Peer behavior, neighborhood standards, local congestion, and status comparisons are all endogenously generated by the market allocation itself. Agents optimize against these social objects, while these social objects must be reproduced by the allocation chosen by agents. Therefore, equilibrium requires consistency not only of aggregate demand and supply, but also of the socially relevant aggregates entering individual preferences.

This consistency requirement, however natural, immediately runs into difficulties once externalities are present. In the canonical continuum economy without externalities (\cite{au66}), one of the principal advantages is that convexity assumptions on individual preferences can be dispensed with. With externalities, as we demonstrate in Section \ref{sec-example}, an equilibrium may fail to exist without convexity. Imposing convex preferences can restore existence under suitable regularity conditions. But doing so would sacrifice precisely one of the most celebrated features of the continuum approach, the ability to accommodate nonconvex preferences in a rigorous, decentralized framework.

Beyond preference nonconvexity, a separate and important source of intrinsic nonconvexity comes from the consumption set itself. When goods such as houses, cars, or school places must be consumed in indivisible integer quantities, the consumption set is inherently nonconvex. This feature cannot be assumed away without altering the economic problem. Replacing the original set by its convex hull introduces fractional consumption, which does not assign an actual bundle of indivisible goods to each individual. Allowing lotteries changes the allocation being studied. Indeed, without imposing further restrictions, the question of equilibrium existence remains largely open even in finite economies with externalities, general preferences and nonconvex consumption sets.\footnote{For a systematic treatment of equilibrium existence in finite-agent economies, see \cite{Florenzano2003}.}

This paper develops a competitive equilibrium framework for large economies with externalities that directly addresses these existence problems. We distinguish between an agent's name and her social type. A name identifies the individual. A type records the attributes through which the individual is socially located. 
The externality faced by an agent is generated by the allocation aggregated at the level of social types. This formulation captures the idea that people usually respond to the behavior of groups rather than to the behavior of every individual in a large society. Our model includes several existing frameworks for large economies and perfect competition.\footnote{See Remark~\ref{rmk-cover} for detailed discussions.}

We demonstrate the role of social information through the illustrative example in Section~\ref{sec-example}. Agents are characterized by location and {educational attainment}. These two characteristics enter social comparisons separately: an agent compares her consumption with cumulative consumption among those whose location and {educational attainment} are both weakly below her own. When exact locations are recorded, individual optimization requires specialization in one of two goods, whereas consistency of the comparison benchmarks and market clearing require balanced consumption on every lower rectangle. These requirements cannot be reconciled, and no equilibrium exists. Now replace exact locations by geographical communities, while retaining  {educational attainment}. Agents in the same community  {with the same educational attainment} then share a social type, but remain distinct individuals. They can consume different goods while keeping the community-level benchmarks balanced. This change restores equilibrium without changing endowments or the utility specification.

\revision{To capture the relative richness of the information conveyed by the agent space and the social type, we impose \emph{social anonymity}: observable social types never exhaust individual identity.\footnote{\revision{It is built on the nowhere equivalence condition introduced by \cite{HSS2017}; see Definition~\ref{def:social-anonymity}.}}}  
\revision{A large economy with externalities thus has two layers of anonymity. Individual agents are negligible for market aggregates, while social anonymity leaves enough room for different individual choices consistent with the same aggregate conditional on social type.} 
Our first main result provides a sharp characterization following this insight. Theorem~\ref{lct-ex1} shows that every large exchange economy  admits an equilibrium if and only if \revision{social anonymity} holds. The sufficiency direction indicates that even if individual demand is nonconvex or affected by the externality, the economy can average choices across agents who share the same social types, thereby generating the required social aggregate. That is, one achieves convexification conditional on social types. The necessity direction implies that when the social-type structure is too rich relative to the agent space, one can construct an economy in which no equilibrium exists. 
\revision{Social anonymity is therefore the minimal structure for equilibrium existence.}

The theorem gives a structural answer to a basic question: when is perfect competition compatible with social interaction?
The message is that the externality needs to aggregate rather than identify. \revision{Competitive markets can accommodate social comparisons, peer effects, and local interactions when the social type function satisfies social anonymity.} They may fail when each agent is able to observe fine details of the allocation, which could be too demanding in a large economy.

Our second main result extends the characterization to generalized large economies with incomplete preferences, production, and both divisible and indivisible goods. These features are important for applications. Some economically important choices are discrete, such as houses, cars, durable facilities, technology adoption, access rights, or local infrastructure. Production shapes the aggregate resource constraint and can make goods available even when their initial supply is zero. Preferences may be incomplete or not representable by a utility function. The framework also permits externalities based on continuous transformations of consumption, including bounded exposure measures and adoption indicators for indivisible goods.\footnote{\cite{Rauh2003} and \cite{YZ2005} study large games in which payoffs depend on summary or nonlinear transformed summary statistics. Remark~\ref{rmk-transformed} gives the corresponding extension here.} Our generalized model contains simultaneously the two natural nonconvexities discussed above: preferences may generate nonconvex demand, and indivisible goods make the feasible consumption set itself nonconvex. An equilibrium allocation must assign actual integer bundles to individuals while reproducing the relevant externality. Despite these difficulties, our Theorem~\ref{lct-ex2} provides a sharp necessary and sufficient characterization of equilibrium existence \revision{in this broader class}, which jointly permits nonconvex preferences and intrinsically nonconvex consumption sets.


Methodologically, we develop novel arguments beyond existing approaches. Our proof replaces the standard aggregate-convexification argument by a conditional-convexification
argument. In the classical continuum model, a continuum of agents convexifies demand at the level of the whole economy. With social externalities, the allocation must also reproduce demand conditional on social types. \revision{Social anonymity is the condition under which this conditional aggregation has the required convexity and compactness properties.} The proof thus connects the economic idea of social anonymity with the theory of conditional expectations of correspondences developed in \cite{HS2018}. For necessity, we provide a sharp result: whenever there is a subeconomy on which social information exhausts individual information, one can construct an economy without any equilibrium, while keeping the number of commodities fixed. 

We demonstrate the usefulness of our results through applications to peer effects, social comparison, and risk sharing. In a peer-effects economy, investment in education or training depends on the investment of classmates, neighbors, cohorts, occupations, or other socially relevant groups.\footnote{See, for example, \cite{Manski1993}, \cite{BD2001}, \cite{Sacerdote2001}, \cite{CPZ2009} and \cite{BBDI2011}.} The peer benchmark is generated by the equilibrium allocation itself. In a social-comparison economy, agents react to the consumption standard of a reference group, as in the ``keeping up with the Joneses'' models.\footnote{See, for example, \cite{Duesenberry1949}, \cite{Abel1990}, \cite{LjungqvistUhlig2000}, and \cite{HopkinsKornienko2004}.} Finally, households can trade state-contingent claims while comparing consumption with endogenous standards in overlapping neighborhoods.\footnote{See, for example, \cite{Arrow1964}, \cite{Wilson1968}, \cite{Townsend1994} and \cite{Abel1990}.} Equilibrium exists even when these comparisons make preferences over state-contingent consumption plans nonconvex. In all three applications, the social object entering preferences is endogenous, while the types defining the reference groups are primitives of the economy.

\paragraph{Related literature.}
This paper contributes to the general equilibrium theory of large economies. Since \cite{au64, au66}, the continuum model has been the standard framework for perfect competition. \cite{hi70, hi74} develops the production and distributional foundations of this approach. \cite{MC77} and \cite{KY81} analyze large economies with indivisible goods.\footnote{There is a  literature on finite markets with indivisible goods under substitution and related conditions; see, for example, \cite{KC82}, \cite{GS99}, and \cite{SY06, SY09}. Pseudomarket approaches study alternative allocation and pricing institutions; see \cite{GPZ24} and \cite{GP26}. These works generally abstract from endogenous consumption externalities.} \cite{Schmeidler69} studies competitive equilibrium with incomplete preferences. Our paper keeps the large-economy formulation but changes the aggregation problem. In the classical model, one only needs to aggregate individual demands at the level of the whole economy. In our model, the allocation also generates the social environment entering preferences, and \revision{the relevant aggregation convexifies demand conditional on social type.}\footnote{\cite{MP02, MP04} study the notion of informational size in economies with asymmetric information. \cite{SY07} study a continuum economy in which each agent's private information affects only a negligible group of agents. These papers and ours share the insight that the economically relevant notion of negligibility depends on the channel through which individuals affect others. 
}

The paper is closely related to the literature on large economies with externalities. \cite{HKH89} study continuum economies with finite coalitions and widespread externalities, and \cite{H99} develops related core-equivalence results. \cite{Noguchi05} gives existence results with convex preferences. \cite{CT05} study the externality mapping with a Lyapunov property, including the case of finitely many fixed reference coalitions.
\cite{NZ06} drop the convexity assumption on preferences by taking a distributional approach.\footnote{When the agent space is modeled by the classical Lebesgue unit interval, as discussed in \cite{SSY2020}, an equilibrium distribution in a large economy  may not be induced by any individual equilibrium allocation. In addition, they show that an equilibrium distribution uniquely corresponds to a (type-symmetric) randomized equilibrium.} They analyze competitive markets with externalities, production and indivisible goods, and provide examples to demonstrate the equilibrium nonexistence issue. 
The recent paper by \cite{ADKU26} works with convex preferences and establishes equilibrium existence in large competitive markets with bads, production, and comprehensive externalities. \revision{We study externalities organized through general social types and identify social anonymity as the precise condition for equilibrium existence.}

The paper is also related to the general-equilibrium literature on welfare and policy under externalities. One line asks when externalities invalidate, preserve, or restore the welfare theorems; see \cite{Ledyard71}, \cite{Osana72}, \cite{Parks91}, \cite{ArrowDasgupta09}, and \cite{LZZ26}. A second line views externalities as missing markets and studies whether efficiency can be decentralized through Arrowian or Lindahl prices for external effects; see \cite{Arrow69}, \cite{Foley70}, \cite{Roberts73}, \cite{KV85},  and \cite{ADL26}. Recently, \cite{AD25} and  \cite{ADKU26} study taxes, quotas, cap-and-trade, and related policy instruments in Arrow--Debreu environments. Before asking whether a competitive allocation is efficient, we ask what kind of social structure makes a competitive allocation possible. Our answer is that large markets require enough anonymity not only in price-taking behavior, but also in the social information through which externalities operate.


The rest of the paper is organized as follows. Section~\ref{sec-example} presents an illustrative example of a social-comparison economy and shows how the richness of social information can determine equilibrium existence. Section~\ref{sec-model}  develops the model of large exchange economies with social types and proves the necessary and sufficient characterization
based on \revision{social anonymity}. It also discusses applications to peer effects, social comparison, and risk sharing. Section~\ref{sec-generalized} extends the result to generalized large economies with incomplete preferences, indivisible goods, and production. Section~\ref{sec-conclusion} concludes. The proofs are collected in the  {Appendix and Supplemental Appendix}.


\section{An illustrative example}\label{sec-example}

This section presents a simple exchange economy with consumption externalities. We first describe an externality structure under which no equilibrium exists, and then show how a coarser organization of the same externality restores equilibrium existence.

Social comparisons often involve more than one characteristic.\footnote{For example, the OECD dimensions are income and wealth, work and job quality, housing, health, knowledge and skills, environmental quality, work–life balance, social connections, civic engagement, safety, and subjective well-being.} A household may compare its consumption with that of people in a particular geographical area  {and with a given level of educational attainment}. These two characteristics need not be combined into a single ranking. In the example below, location and  {educational attainment} enter the comparison directly. The difference between the two cases is whether geographical information identifies an exact location or only a broader community.

Each individual is indexed by a pair $(v,w)$. The first coordinate $v\in V=[0,1]$ denotes her location along a geographical axis, and the second coordinate $w\in W=[0,1]$ denotes her  {educational attainment, modeled as a continuous characteristic with larger values indicating more schooling}. The population is $I=V\times W$, endowed with the completed Lebesgue $\sigma$-algebra $\cI$ and Lebesgue probability measure $\lambda$. The social type space is $T=[0,1]^2$. A social type \revision{function} $\alpha:I\to T$ records the location and  {educational} information used in comparisons. The order on $T$ is coordinatewise: a comparison includes agents whose recorded location and  {educational attainment} are both weakly below those of the individual.

Each agent is endowed with one unit of each of two divisible goods, $e(v,w)=(1,1)$. There is no production. A consumption allocation is a measurable and integrable function $f=(f^1,f^2):I\to\mathbb R_+^2$. For $k=1,2$, define the comparison index at social type $s$ by
\[
\eta_k(s;f)=\int_{\{(t,j)\in I:\,\alpha(t,j)\le s\}}f^k(t,j)\,d\lambda(t,j).
\]
Thus, $\eta_k$ records cumulative consumption of good $k$ in the comparison group. It is an aggregate, rather than an average, and is nondecreasing in each coordinate of $s$.

The utility of agent $(v,w)$ from the bundle $(x^1,x^2)$ is
\[
u_{(v,w)}(x^1,x^2;f)
=\left(\frac{x^1}{1+\eta_1(\alpha(v,w);f)}\right)^{\frac32}
+\left(\frac{x^2}{1+\eta_2(\alpha(v,w);f)}\right)^{\frac32}.
\]
Greater consumption of a good in the comparison group reduces the marginal value of own consumption of that good. For fixed comparison indices, utility is continuous, strictly increasing in each good, and strictly convex. Individual demand therefore favors specialization: an agent spends her entire income on one of the two goods.

\paragraph{Case~I:} \textbf{Comparisons based on exact location and  {educational attainment}.}

Let $\alpha(v,w)=(v,w)$. The comparison group of agent $(v,w)$ is the lower rectangle $[0,v]\times[0,w]$. Hence
\[
\eta_k(\alpha(v,w);f)=\int_0^v\int_0^w f^k(t,j)\,dj\,dt,
\qquad k=1,2.
\]
Both coordinates of an individual's social type are observed exactly. 

\begin{fact}\label{fact-1}
The economy with $\alpha(v,w)=(v,w)$ has no equilibrium.
\end{fact}

To see the conflict, suppose that $(f,p)$ is an equilibrium. Strict monotonicity requires $p^1,p^2>0$; normalize $p^1+p^2=1$. Every agent has income one. Strict convexity then implies that almost every agent chooses either $(1/p^1,0)$ or $(0,1/p^2)$.  {Spending the entire income on good $k$ gives utility}
\[
\bigl\{p^k[1+\eta_k(\alpha(v,w);f)]\bigr\}^{-\frac32}.
\]
Thus, the agent chooses the first good when its price adjusted by the comparison index is lower, and the second good when the reverse inequality holds.

Write $F_k(v,w)=\eta_k(\alpha(v,w);f)$ and consider the difference
\[
H(v,w)=p^1[1+F_1(v,w)]-p^2[1+F_2(v,w)].
\]
With $c=p^1-p^2$ and $q=p^1f^1-p^2f^2$, this becomes
\[
H(v,w)=c+\int_0^v\int_0^w q(t,j)\,dj\,dt.
\]
 {Almost every agent spends her entire income on one good, so $q=1$
if she buys good 1 and $q=-1$ if she buys good 2.
When $H(v,w)>0$, she strictly prefers to buy only good 2;
when $H(v,w)<0$, she strictly prefers to buy only good 1.
Thus $q=-1$ almost everywhere on $\{H>0\}$ and $q=1$ almost
everywhere on $\{H<0\}$. Lemma~\ref{lem:rectangle} in Appendix~\ref{app:proof-theorem1} then gives}
\[
|H(v,w)|\le |p^1-p^2|
\qquad\text{for every }(v,w)\in I.
\]

Market clearing requires $F_1(1,1)=F_2(1,1)=1$. At the upper corner of the square, therefore, $H(1,1)=2(p^1-p^2)$. The bound forces $p^1=p^2$. It then also forces $H(v,w)=0$ on every lower rectangle, so $F_1(v,w)=F_2(v,w)$ for every $(v,w)$. Taking the mixed derivative of these cumulative integrals gives $f^1=f^2$ almost everywhere. This contradicts the requirement that every agent specialize in one good. No equilibrium exists.


Even though the population is atomless, equilibrium would require almost every agent to consume equal amounts of the two goods.
This is incompatible with individual optimality, which requires almost every agent to spend her entire income on one good.

\paragraph{Case~II:} \textbf{Community-level comparisons.}
Suppose now that agents do not condition their comparisons on exact locations. Partition $V$ into finitely many geographical communities with endpoints $0=a_0<a_1<\cdots<a_M=1$. 
Write $C_m=[a_{m-1},a_m)$, with the last interval including $1$. For $v\in C_m$, let
\[
\alpha_M(v,w)=\left(\frac{a_{m-1}+a_m}{2},w\right).
\]
This indicator replaces an individual's location by the midpoint of her community and retains her  {educational attainment}.

Use the same utility and comparison formulas with $\alpha_M$ in place of $\alpha$. An agent in community $m$ with  {educational attainment} $w$ compares consumption with agents in communities $1,\ldots,m$ whose  {educational attainment does} not exceed $w$. Her comparison indices are therefore
\[
\eta_k(\alpha_M(v,w);f)
=\int_0^{a_m}\int_0^w f^k(t,j)\,dj\,dt,
\qquad k=1,2.
\]
The reference group now includes the whole geographical community, rather than stopping at the individual's own location.

\begin{fact}\label{fact-2}
For every finite community partition, the economy with community-level comparisons admits an equilibrium.
\end{fact}

Take $(p^1,p^2)=(1/2,1/2)$. Within each community, assign the first good to the first half of its locations and the second good to the remaining half:
\[
f(v,w)=
\begin{cases}
(2,0),&a_{m-1}\le v<(a_{m-1}+a_m)/2,\\
(0,2),&(a_{m-1}+a_m)/2\le v<a_m,
\end{cases}
\qquad v\in C_m.
\]
 {For $v=1$, set $f(1,w)=(0,2)$.}
At every  {level of educational attainment}, each community has equal average consumption of the two goods. Consequently,
\[
\eta_1(\alpha_M(v,w);f)
=\eta_2(\alpha_M(v,w);f)=a_mw.
\]
 {At these prices, buying only good 1 and buying only good 2 are both
optimal. The proposed allocation assigns one of these bundles to
each agent and clears both markets.}

\revision{The contrast highlights the role of the social type function relative to the agent space.} In Case~I, an agent's type records both coordinates exactly. In Case~II, agents in the same community  {with the same educational attainment} share a social type, while their individual locations remain distinct. They can receive different bundles without changing the community-level comparison indices. This possibility reconciles individual specialization with balanced social aggregates. \revision{In the next section, we formalize this requirement as social anonymity.}

\section{Large exchange economies}\label{sec-model}

In this section, we present the model of large exchange economies with externalities. The economy consists of a continuum of agents and $l$ divisible goods. Each agent is endowed with an initial bundle. We extend the standard model by introducing social types, which may represent location, occupation, or other observable characteristics. An agent's payoff depends on both her consumption bundle and the allocation aggregated by social type. As demonstrated in the previous section, an equilibrium may fail to exist. \revision{We shall show how social anonymity characterizes equilibrium existence in economies satisfying the assumptions below.}

\subsection{Setup}\label{subsec-setup}

Formally, a large exchange economy is as follows.

\begin{enumerate}
\item The agent space is an atomless and complete probability space $(I,\mathcal I,\lambda)$.

\item There are $l\geq1$ divisible goods, with commodity space $\mathbb R_+^l$. The endowment mapping $e:I\to\mathbb R_+^l$ is integrable and satisfies $\int_I e\,d\lambda\gg0$.\footnote{For vectors $x$ and $y$, $x\gg y$ means that $x^j>y^j$ in every coordinate, and $x\geq y$ means that $x^j\geq y^j$ in every coordinate. For a finite-dimensional vector $x$, write
$\|x\|_1=\sum_j|x^j|$.}

\item An allocation is an $\mathcal I$-measurable, integrable function $f:I\to\mathbb R_+^l$.

\item Let $T$ be the space of possible social types. It is a complete, separable metrizable space, endowed with its Borel $\sigma$-algebra $\mathcal B(T)$. The measurable social type  function $\alpha:I\to T$ assigns a type $\alpha(i)$ to each agent $i$.

\item \revision{Let $\mathcal G\subseteq\mathcal I$ denote the $\sigma$-algebra generated by the social type function $\alpha$. It is countably generated because $T$ is separable and metrizable. Given an allocation $f$, the externality is its conditional expectation given $\alpha$, denoted by $\mathbb E(f\mid\alpha)$.}\footnote{\revision{Conditional expectations and all $L_1$ functions below are identified up to equality almost everywhere.} }

\revision{Let $\Gamma=L_1((I,\mathcal G,\lambda);\mathbb R_+^l)$ be the set of possible externalities, endowed with the weak topology inherited from $L_1((I,\mathcal G,\lambda);\mathbb R^l)$.} 
Throughout the paper, a bounded sequence in $\Gamma$ means a sequence bounded in the $L_1$ norm.\footnote{\revision{For $\gamma\in L_1((I,\mathcal G,\lambda);\mathbb R^l)$,} 
\[
\|\gamma\|_{L_1}
:=
\int_I\|\gamma(i)\|_1\,d\lambda(i)
=
\sum_{j=1}^l\int_I|\gamma^j(i)|\,d\lambda(i).
\]}

\item Each agent $i$ has a utility function $u_i:\mathbb R_+^l\times\Gamma\to\mathbb R$, jointly continuous in consumption and the externality. For each $(x,\gamma)$, the mapping $i\mapsto u_i(x,\gamma)$ is $\mathcal I$-measurable. For every $i$ and $\gamma$, $u_i(\cdot,\gamma)$ is strictly monotone in own consumption.\footnote{That is, $x\geq y$ and $x\neq y$ imply $u_i(x,\gamma)>u_i(y,\gamma)$.} Given an allocation $f$, agent $i$'s payoff is $u_i(f(i),\mathbb E(f\mid\alpha))$.

\item Let $\wp$ be the space of continuous functions on $\mathbb R_+^l$, endowed with the compact-open topology, and let $\wp_+$ be its subset of strictly monotone functions.\footnote{The compact-open topology on this space is the topology of uniform convergence on compact consumption sets.} For each $\gamma\in\Gamma$, define $\Phi_\gamma(i)=u_i(\cdot,\gamma)$ and let $\mu_\gamma=\lambda\circ\Phi_\gamma^{-1}$ be the induced distribution on $\wp$. The measurability in (6) makes $\Phi_\gamma$ measurable.

For every bounded sequence $\{\gamma_n\}_{n\geq1}\subset\Gamma$, the sequence $\{\mu_{\gamma_n}\}_{n\geq1}$ is uniformly tight: for every $\epsilon>0$, there is a compact set $K_\epsilon\subset\wp_+$ such that $\mu_{\gamma_n}(K_\epsilon)\geq1-\epsilon$ for every $n$.
\end{enumerate}

A large economy is a pair $(\mathcal E,\alpha)$, where $\mathcal E(i)=(u_i,e(i))$.

\begin{rmk}\label{rmk-externality}
Individual utility may be influenced by others' consumption through status seeking, conformism, or altruism. It may nevertheless be implausible that agents in a large economy respond to the precise consumption bundle of every individual, as it would impose unrealistic requirements or cognitive loads on the agents. Our formulation reflects the idea that agents categorize others through observable types, such as social position, occupation, or geographic location. The conditional expectation $\mathbb E(f\mid\alpha)$ summarizes consumption at the level of these types. \revision{An agent may respond to consumption averages over several sets of social types, or to weighted averages across types, without responding to the consumption of each individual separately.}
\end{rmk}

\begin{rmk}\label{rmk-cover}
Individual utilities may vary with the social environment. Along any
norm-bounded sequence of environments, condition~(7) allows us to
choose a single compact family of strictly increasing utility functions
that contains the utilities of nearly all agents in each environment.
The share of exceptions can be made arbitrarily small, and the agents
concerned may differ across environments. The condition is satisfied in several familiar settings.
\begin{enumerate}
\item[(a)] It encompasses the standard large exchange economy of \cite{au64,au66}, where $u_i$ is independent of $\gamma$. 

\item[(b)] More generally, suppose $u_i(x,\gamma)=v_i(x,z(\gamma))$, where $z$ is continuous, takes values in a compact metric space $Z$, and each $v_i$ is continuous in $(x,z)$ and strictly monotone in $x$ for every $z$. Assume also that $i\mapsto v_i(x,z)$ is measurable for every $(x,z)$. A compact family of the functions $v_i$ covering all but a small fraction of agents yields a compact family of their consumption sections, uniformly over $z\in Z$. The same argument applies if $z(\gamma)$ ranges over a compact set whenever $\gamma$ ranges over a norm-bounded set. This includes externalities described by finitely many group aggregates, such as $(\int_{I_1}f\,d\lambda,\ldots,\int_{I_n}f\,d\lambda)$ for a $\mathcal G$-measurable partition $I_1,\ldots,I_n$, as in \cite{HKH89}; see also \cite{H99}.

\item[(c)] Suppose agents have measurable preference parameters $\theta_i\in\Theta$, where $\Theta$ is compact, and externalities enter through indices $z_i(\gamma)\in Z$, where $Z$ is compact. If $u_i(x,\gamma)=v(x,\theta_i,z_i(\gamma))$, with $v$ continuous and strictly monotone in $x$ for every $(\theta,z)$, then
\[
K=\{v(\cdot,\theta,z): (\theta,z)\in\Theta\times Z\}
\]
is a compact subset of $\wp_+$. Thus (7) holds. As in (b), the compact set of indices may depend on a bound for $\|\gamma\|_{L_1}$. This formulation accommodates the description of agents' characteristics as in \cite{hi74}.
\end{enumerate}

For the existence result, uniform tightness can be replaced
by the following ordinal condition. Let $q^j$ be the $j$th coordinate unit vector. For every $M,R>0$ and $\epsilon>0$, suppose that there is $\eta\in(0,1)$ such that, for every $\gamma$ with $\|\gamma\|_{L_1}\le M$, there is a measurable set $I_\gamma\subseteq I$ with $\lambda(I_\gamma)\ge1-\epsilon$ on which
\begin{equation}\label{eq:ordinal-stability}
u_i((1-\eta)x+q^j,\gamma)>u_i(x,\gamma)
\quad\text{for all }\|x\|_1\le R,\ x\in\mathbb R_+^l,
\quad j=1,\ldots,l.
\end{equation}
Thus, on bounded consumption sets, an additional unit of any good compensates for a sufficiently small proportional reduction in the bundle, uniformly over bounded social environments and outside an arbitrarily small set of agents. The exceptional set may depend on the social environment. This condition depends only on preference orderings and is unchanged by strictly increasing transformations of individual utilities. Uniform tightness implies \eqref{eq:ordinal-stability}: continuity and strict monotonicity give a common $\eta$ on each compact family of utilities, and failure of the stated uniformity would produce a bounded sequence of externalities violating uniform tightness. We explain how to make the corresponding modification for the proof of Theorem~\ref{lct-ex1} in the Appendix.

\end{rmk}

Below, we illustrate the structure of the externality through two examples.

\begin{exam}
Suppose the agent space is the completed Lebesgue unit interval. If the social type function $\alpha:[0,1]\to\mathbb R$ is strictly monotone, it distinguishes every agent. \revision{Consequently, $\mathbb E(f\mid\alpha)=f$ almost everywhere.}

Alternatively, let $\alpha(i)=L$ for $i\in[0,1/2)$ and $\alpha(i)=H$ otherwise. The social-type information distinguishes only the two groups. The conditional expectation is their average consumption: $\mathbb E(f\mid\alpha)$ equals $2\int_0^{1/2}f\,d\lambda$ on the first group and $2\int_{1/2}^1f\,d\lambda$ on the second.
\end{exam}

\begin{exam}[Revisiting the example in Section~\ref{sec-example}]\label{ex:rectangle}
The comparison indices $\eta_1$ and $\eta_2$ are functions of $\mathbb E(f\mid\alpha)$ in Case~I and of $\mathbb E(f\mid\alpha_M)$ in Case~II. Thus, the framework above covers the example in Section~\ref{sec-example}.
\begin{itemize}
\item \revision{In Case~I, $\alpha(v,w)=(v,w)$ identifies the agent, so $\mathbb E(f\mid\alpha)=f$ almost everywhere. The lower rectangle $R_{v,w}=[0,v]\times[0,w]$ is determined by social type, and}
\[
\eta_k(\alpha(v,w);f)
=\int_{R_{v,w}}f^k\,d\lambda
=\int_{R_{v,w}}\mathbb E(f^k\mid\alpha)\,d\lambda,
\qquad k=1,2.
\]
\item \revision{In Case~II, the social type function $\alpha_M$ records community and educational attainment.} For almost every $(v,w)$ with $v\in C_m$, conditional expectation averages over locations within the community:
\[
\mathbb E(f\mid\alpha_M)(v,w)
=\frac{1}{a_m-a_{m-1}}\int_{a_{m-1}}^{a_m}f(t,w)\,dt.
\]
The comparison set of an agent in community $m$  {with educational attainment} $w$ is $R_{m,w}=(C_1\cup\cdots\cup C_m)\times[0,w]$. \revision{This set is determined by social type and equals $[0,a_m]\times[0,w]$ (except the boundary).} Hence
\[
\eta_k(\alpha_M(v,w);f)
=\int_{R_{m,w}}f^k\,d\lambda
=\int_{R_{m,w}}\mathbb E(f^k\mid\alpha_M)\,d\lambda,
\qquad k=1,2.
\]
Thus, averaging within communities preserves every comparison index, while allowing agents with the same social type to choose different bundles.
\end{itemize}

The other assumptions are satisfied as well. Integration over a fixed comparison rectangle is a weakly continuous functional on $\Gamma$. The utility of each agent is therefore jointly continuous in own consumption and the externality, and it is strictly monotone in own consumption. The utility assignment is measurable. Indeed, when rectangle endpoints converge, their symmetric differences have measure tending to zero. Uniform integrability on every weakly compact subset of $\Gamma$ then gives uniform convergence of their integrals on that subset. Thus, the assignment is continuous in $(v,w)$ in Case~I and continuous in $w$ within each community in Case~II, under the compact-open topology.

Finally, let $(\gamma_n)$ be a norm-bounded sequence of externalities, and choose $b<\infty$ bounding $\|\gamma_n\|_{L_1}$ for every $n$. Every comparison integral lies in $[0,b]$. All conditional utilities therefore belong to the compact family
\[
\left\{x\longmapsto b_1(x^1)^{3/2}+b_2(x^2)^{3/2}:
(b_1,b_2)\in[(1+b)^{-3/2},1]^2\right\}.
\]
Every member is strictly monotone, which verifies uniform tightness.
\end{exam}

\begin{defn}
An allocation $f$, together with a nonnegative, nonzero price vector $p$, is an equilibrium of $(\mathcal E,\alpha)$ if the following conditions hold.
\begin{enumerate}
\item Utility maximization: for almost every $i\in I$,
\[
f(i)\in D(i,p,\mathbb E(f\mid\alpha)),\qquad
D(i,p,\gamma)=\argmax_{x\in B(i,p)}u_i(x,\gamma),
\]
where $B(i,p)=\{x\in\mathbb R_+^l:p\cdot x\leq p\cdot e(i)\}$.
\item Market clearing: $\int_I f\,d\lambda=\int_I e\,d\lambda$.
\end{enumerate}
\end{defn}

\subsection{Equilibrium existence}\label{subsec-existence}

The illustrative example in Section~\ref{sec-example} holds endowments and the utility specification fixed while changing the social information through which consumption is compared. In both cases, agents care about others' consumption. \revision{The difference lies in whether the social type function leaves room for different individual choices while preserving the comparison benchmarks.}

In Case~I, the pair of location and  {educational attainment} identifies the agent, up to null sets. Individual optimization requires each agent to specialize in one good. Yet the rectangular comparison structure, together with market clearing, forces cumulative consumption of the two goods to agree on every lower rectangle. Since these rectangles determine the allocation, the two requirements are incompatible.

In Case~II, location is recorded only at the community level. For each  {level of educational attainment}, the population within a community can be divided between the two consumption choices. This division keeps average consumption equal to $(1,1)$  {among agents in the same community with the same educational attainment} and therefore reproduces the comparison benchmarks that support the choices. The relevant richness is in the population within social types; it need not come from differences in preferences or endowments.

\revision{We formalize this distinction as \emph{social anonymity}, using the nowhere equivalence condition introduced by \cite{HSS2017}.} 
For a positive-measure set \(D\in\mathcal I\), define the restricted \(\sigma\)-algebras $\mathcal I^D =    \{D\cap A:A\in\mathcal I\}$ and $\mathcal G^D = \{D\cap A:A\in\mathcal G\}$. \revision{The $\sigma$-algebra $\mathcal I$ is nowhere equivalent to $\mathcal G$ if, for every non-negligible set $D\in\mathcal I$, there is an $\mathcal I$-measurable subset $D_0\subseteq D$ such that $\lambda(D_0\triangle D_1)>0$ for every $D_1\in\mathcal G^D$.} 


\begin{defn}[\revision{Social anonymity}]\label{def:social-anonymity}
\revision{A large economy with social type function $\alpha$ is \emph{socially anonymous} if $\mathcal I$ is nowhere equivalent to $\mathcal G$. We also say that $\alpha$ satisfies \emph{social anonymity} relative to the agent space $(I,\mathcal I,\lambda)$.}
\end{defn}

\revision{Social anonymity is a condition on the agent space and the social type function. The function $\alpha$ records the characteristics through which externalities operate, such as community, location, wealth rank, or occupation.}
\revision{Social anonymity requires that these characteristics leave measurable individual distinctions on every positive-measure part of the population.\footnote{\revision{For example, a constant social type function satisfies social anonymity on an atomless agent space.}} Thus, even after restricting attention to a subeconomy, social types cannot identify all its measurable subsets of agents. In Section~\ref{sec-example}, social anonymity fails in Case~I and holds in Case~II, where individual location continues to vary after community and educational attainment have been recorded.}

\revision{The following theorem characterizes equilibrium existence in terms of social anonymity.}

\begin{thm}\label{lct-ex1}
\revision{Fix an integer $l\geq2$, an agent space $(I,\mathcal I,\lambda)$, and a social type function $\alpha:I\to T$. The following statements are equivalent.}
\begin{enumerate}
\item[i.] \revision{The social type function satisfies social anonymity.}
\item[ii.] Every $l$-good large exchange economy admits an equilibrium.
\end{enumerate}
\end{thm}

\begin{rmk}
The sufficiency part says that idiosyncratic variation within social types is enough to restore equilibrium existence. The proof uses the conditional-expectation correspondence associated with individual demand. \revision{Under social anonymity, this correspondence has the convexity and compactness needed for a fixed-point argument, even when individual demand is nonconvex.} Different optimal bundles can be assigned within the population while preserving the required conditional aggregate.

\revision{For necessity, failure of social anonymity identifies a positive-measure subeconomy on which social types exhaust individual information. The rectangular example can be placed in that subeconomy without changing the given social type function $\alpha$.} Individual optimization requires specialization, whereas consistency of rectangular consumption totals requires equal consumption of the two goods almost everywhere. The construction uses two goods and extends to any fixed $l\geq2$ by combining the remaining goods into a single consumption index. It does not require a sequence of economies with an increasing number of goods.

Therefore, Theorem~\ref{lct-ex1} gives a sharp characterization of equilibrium existence, in the sense that it provides the minimal structural condition on the relation between agents’ identities and social types.

\end{rmk}

\begin{rmk}\label{rmk-tightness-essential}
\revision{Social anonymity and uniform tightness address different difficulties.} The former permits individual choices to vary within social types while preserving the externality. The latter prevents preferences from degenerating along bounded sequences of social environments. In  {Supplemental Appendix~\ref{app:tightness-proof}}, we give a two-good economy satisfying all the assumptions of Section~\ref{subsec-setup} except uniform tightness, with \revision{social anonymity}, identical strictly positive endowments, and linear strictly monotone utilities, but no equilibrium. Along resource-feasible allocations, consumption of the first good becomes concentrated among low social types, making almost all agents increasingly reluctant to give up the second good. 
The example also violates the ordinal condition in Remark~\ref{rmk-cover}.
\end{rmk}

\revision{When payoffs do not depend on externalities, the social type function $\alpha$ can be taken to be constant. Atomlessness then implies social anonymity.} Theorem~\ref{lct-ex1} implies the equilibrium existence result in the canonical continuum economy without externalities as in \cite{au66}.

\begin{cro}
A large exchange economy without externalities has an equilibrium.
\end{cro}

\subsection{Applications}\label{subsec-applications}

In the following, we provide three economic applications to illustrate the framework. In each application, the social benchmark is determined by the market allocation, while the social types that define the reference groups are given as part of the economy.

\paragraph{Peer effects in education and human-capital investment}
\ 

We first consider an economy with peer effects in education. The example is motivated by the literature on social interactions and peer effects; see, for example, \citet{Manski1993}, \citet{BD2001}, \citet{Sacerdote2001} and \citet{CPZ2009}. An individual's return to an activity may depend on the behavior of her peers. In a competitive economy, both individual investment and the peer behavior that affects its return are determined by the market allocation.

There are two divisible goods. Good $0$ is private consumption, and good $1$ is education or training services. A bundle is written as $x=(c,h)\in\mathbb R^2_+$, where $h$ is the amount of services purchased for human-capital investment. The endowment $e:I\to\mathbb R^2_+$ is integrable, with $\int_I e\,d\lambda\gg0$. Its second coordinate represents claims to a fixed supply of training services. Thus, it is the input into human-capital investment that is traded, rather than the resulting stock of human capital.

\revision{Each agent has a social type $\alpha(i)\in T$, where $T$ is a compact metric space.} The social type may describe a school, neighborhood, cohort, or occupation. For example, take $I=T\times[0,1]$ endowed with $\nu \otimes \mu$, where $\nu$ is a probability measure on $T$ and $\mu$ is the uniform distribution on $[0,1]$. Let $\alpha(t,r)=t$. The coordinate $r$ indexes students or households with the same social type. \revision{With this product population, the social type function satisfies social anonymity.}

For $\gamma=(\gamma^0,\gamma^1)\in\Gamma$, define agent $i$'s peer-exposure index by
\[
H_i(\gamma)=\int_I k(\alpha(i),\alpha(i'))\gamma^1(i')\,d\lambda(i'),
\]
where $k:T\times T\to\mathbb R_+$ is bounded and Borel measurable. The weight $k(t,t')$ measures the exposure of an agent with type $t$ to investment by agents with social type $t'$. It can place greater weight on agents in the same school or nearby neighborhoods. When the weights integrate to one, $H_i$ is a weighted average of peer investment. \revision{For fixed $i$, the weight $k(\alpha(i),\alpha(i'))$ is a bounded measurable function of $\alpha(i')$, so}
\[
H_i(\mathbb E(f\mid\alpha))
=\int_I k(\alpha(i),\alpha(i'))f^1(i')\,d\lambda(i').
\]
Thus, the conditional expectation contains precisely the information needed to compute peer exposure.

Agent $i$ has a preference parameter $\theta_i\in\Theta$, where $\Theta$ is compact metric and $i\mapsto\theta_i$ is measurable. This parameter may describe ability, family background, or the private return to training. Let $A,B:\mathbb R_+\times\Theta\times T\to\mathbb R$ and $Q:\mathbb R_+\times\Theta\times T\to\mathbb R_+$ be continuous. Suppose that $A(\cdot,\theta,t)$ and $B(\cdot,\theta,t)$ are strictly increasing, and that $Q(\cdot,\theta,t)$ is nondecreasing. Let $\beta:\mathbb R_+\times\Theta\times T\to\mathbb R_+$ also be continuous. Agent $i$'s utility is
\[
u_i(c,h,\gamma)
=A(c,\theta_i,\alpha(i))+B(h,\theta_i,\alpha(i))
+\beta(H_i(\gamma),\theta_i,\alpha(i))Q(h,\theta_i,\alpha(i)).
\]
The first two terms describe the utility from private consumption and the private return to training. The last term allows that return to depend on peer investment. If $\beta$ is nondecreasing in its first argument, higher peer investment raises the incremental return to training. This gives a complementarity between individual investment and the investment of the reference group. No concavity assumption is imposed on the return to training.

These preferences satisfy the assumptions of Section~\ref{subsec-setup}. For each agent, $H_i$ is weakly continuous, and the utility assignment is measurable. If $\|\gamma\|_{L_1} \le M$, then $0\le H_i(\gamma)\le\|k\|_\infty M$. The conditional utilities therefore belong to the continuous image of the compact parameter set $\Theta\times T\times[0,\|k\|_\infty M]$ in the compact-open topology. Every utility in this image is strictly monotone. Hence, the uniform-tightness condition holds.

\revision{By Theorem~\ref{lct-ex1}, the economy admits an equilibrium whenever social anonymity holds.} 
Economically, peer effects create an endogenous feedback loop: each agent’s demand for human-capital investment depends on peer-group investment, while peer-group investment
is generated by the equilibrium allocation itself. The model does not require agents to observe the investment choice of every individual. Instead, they react to social information
summarized by social types: schools, neighborhoods, cohorts, occupations, or network positions. The result permits nonconcave returns and does not require peer effects to be small.

\paragraph{Keeping up with the Joneses}
\ 

Next, we consider social comparisons that change the relative value of different consumption goods. The example is motivated by models of relative income, status, and ``keeping up with the Joneses'' preferences; see, for example, \cite{Duesenberry1949}, \cite{Abel1990}, \cite{LjungqvistUhlig2000}, and \cite{HopkinsKornienko2004}. Households compare their consumption with that of a reference group, and the comparison may affect the value of each good differently. This example demonstrates that our result is applicable to a broad class of models with the coefficient formulation of social pressure.

There are $l\geq2$ divisible goods, with integrable endowments satisfying $\int_I e\,d\lambda\gg0$. For a concrete social structure, let $I=[0,1]^2$. Write an agent as $i=(v,w)$, where $v$ is her position in a given social ordering and $w$ indexes households at that position. \revision{Set $\alpha(v,w)=v$.} The ordering may reflect occupational or social background; it is not determined by the agent's equilibrium consumption.

The reference group of agent $i=(v,w)$ consists of households weakly below her position: $L_i=[0,v]\times[0,1]$. For $v>0$, define the good-specific comparison index by the average consumption of this group,
\[
\xi_i^j(\gamma)=\frac1v\int_{L_i}\gamma^j(i')\,d\lambda(i'),
\qquad j=1,\ldots,l.
\]
Set $\xi_i^j(\gamma)=0$ when $v=0$, which concerns only a null set of agents. \revision{Since membership in $L_i$ is determined by social type,} the index $\xi_i^j(\mathbb E(f\mid\alpha))$ is the actual average consumption of good $j$ in the reference group. Normalizing by the group's measure separates its consumption standard from its size.

Let $\theta_i\in\Theta$ be measurably assigned, with $\Theta$ compact metric. For each good $j$, let $v_j:\mathbb R_+\times\Theta\to\mathbb R$ be continuous and strictly increasing in its first argument. Let $a_j:\mathbb R_+\to[\underline a_j,\overline a_j]$ be continuous, where $0<\underline a_j\le\overline a_j<\infty$. Agent $i$'s utility is
\[
u_i(x,\gamma)=\sum_{j=1}^l a_j(\xi_i^j(\gamma))v_j(x^j,\theta_i).
\]
If $a_j$ is increasing, higher reference-group consumption raises the utility gain from additional consumption of good $j$. This captures a keeping-up motive. A decreasing $a_j$ instead allows reference-group consumption to reduce the value of additional own consumption. The specification concerns the effect of a social benchmark on relative valuations; it does not require a household to know its rank in the distribution of consumption.

Each comparison index is weakly continuous for a fixed agent. The factor $1/v$ is not uniformly bounded across agents, but this causes no difficulty for uniform tightness: all conditional utilities lie in the compact family
\[
\left\{x\mapsto\sum_{j=1}^l b_jv_j(x^j,\theta):
\theta\in\Theta,\ b_j\in[\underline a_j,\overline a_j]\right\}
\]
of strictly monotone functions. The utility assignment is measurable. \revision{In this product population, the unrecorded household coordinate $w$ supplies the individual variation required for social anonymity, and Theorem~\ref{lct-ex1} gives an equilibrium.}

\paragraph{Risk sharing with local social benchmarks}
\ 

Our formulation also permits a risk-sharing application with overlapping local reference groups. The allocation of risk is a classical question in the theory of contingent markets and risk sharing; see \citet{Arrow1964} and \citet{Wilson1968}. Consumption insurance within local communities is also central to the empirical analysis of \citet{Townsend1994}. Here households trade in complete contingent markets, but evaluate consumption relative to endogenous neighborhood standards. This connects the risk-sharing problem to consumption externalities such as those studied by \citet{Abel1990}. We ask whether competitive equilibrium exists when gain--loss preferences around these standards need not be convex.

There are $l\ge2$ states with probabilities $\pi^j>0$ and $\sum \pi^j = 1$. Let $I=[0,1]^2$, and write a household as $(v,w)$. The coordinate $v$ is its location and $w$ indexes households and their preference characteristics at that location. \revision{Social types record location: $\alpha(v,w)=v$.} An integrable endowment $e(v,w)\in\mathbb R_+^l$ specifies the household's resources in each state, with $\int_I e\,d\lambda\gg0$. At state prices $p$, the household can choose any consumption plan $x$ satisfying $p\cdot x\le p\cdot e(v,w)$.

Fix $0<\delta\le1$. The neighborhood of location $v$ is $N_\delta(v)=[v-\delta,v+\delta]\cap[0,1]$. Writing $|N_\delta(v)|$ for its length, define the state-$j$ consumption standard by
\[
R_v^j(\gamma)=\frac{1}{|N_\delta(v)|}
\int_{N_\delta(v)}\int_0^1\gamma^j(t,r)\,dr\,dt.
\]
This is the average consumption in the neighborhood. \revision{Since membership in $N_\delta(v)\times[0,1]$ is determined by social type,} the same average is obtained from $f$ and from $\mathbb E(f\mid\alpha)$. Households respond to local consumption standards without observing the consumption of individual neighbors.

Let $b:[0,1]\to[\underline b,\overline b]$ be continuous, with $0<\underline b\le\overline b<\infty$. The parameter $b(w)$ measures the household's sensitivity to social comparisons. Consider the gain--loss function
\[
m(z)=
\begin{cases}
z^\rho,&z\ge0,\\
-\kappa(-z)^\rho,&z<0,
\end{cases}
\qquad 0<\rho<1,\quad \kappa>1,
\]
and preferences
\[
u_{(v,w)}(x,\gamma)=\sum_{j=1}^l\pi^j
\left[x^j+b(w)m\!\left(x^j-R_v^j(\gamma)\right)\right].
\]
The linear term values material consumption. The comparison term assigns greater weight to a loss than to an equally sized gain, and exhibits diminishing sensitivity on each side of the reference point, as in \citet{KT1979}. We retain ordinary probability weights and let the reference point depend on the consumption of the neighborhood. In the loss region, utility from state-$j$ consumption is convex. Consequently, preferences over consumption plans can be nonconvex, and the usual concavity argument for competitive risk sharing does not apply.

Nevertheless, the assumptions of Theorem~\ref{lct-ex1} are satisfied. For each location, the reference point is a bounded linear functional of the externality. Since $|N_\delta(v)|\ge\delta$, a bound $\|\gamma\|_{L_1} \le M$ implies $0\le R_v^j(\gamma)\le M/\delta$ for every location and state. The conditional utilities thus belong to the compact family indexed by $b\in[\underline b,\overline b]$ and reference vectors in $[0,M/\delta]^l$. All these utilities are continuous and strictly increasing in own consumption. \revision{Their assignment to households is measurable, and the product population satisfies social anonymity because the atomless household coordinate $w$ is not recorded by $\alpha$.}

Therefore, there is a competitive equilibrium $(f,p)$ in which households choose their preferred state-contingent consumption plans, claims clear in every state,
\[
\int_I f^j\,d\lambda=\int_I e^j\,d\lambda,\qquad j=1,\ldots,l,
\]
and neighborhood standards are generated by those same plans. The result permits nonconvex attitudes toward risk and imposes no smallness restriction on the strength of social comparisons. It establishes existence, without requiring common risk preferences or implying uniqueness or efficient risk sharing.

\section{Generalized large economies}\label{sec-generalized}

The exchange economy in Section~\ref{sec-model} isolates the role of social types in an environment with utility representations, divisible goods, and fixed aggregate resources. Many applications require additional features. Households may have incomplete preferences; commodities such as houses, cars, or durable facilities may be indivisible; and resources may be shaped by production. These features introduce new difficulties. In this section, we develop a general framework that extends the model along three dimensions: (i) agents have incomplete preferences; (ii) there are indivisible goods; and (iii) there is a production sector. \revision{We shall show that social anonymity is again necessary and sufficient for every economy in this broader class to admit an equilibrium.}

\subsection{Setup}\label{glc}

The generalized economy is described as follows.
\begin{enumerate}
\item The agent space is an atomless and complete probability space $(I,\mathcal I,\lambda)$.

\item There are $l\geq1$ divisible goods and $m\geq0$ indivisible goods, with commodity space $X=\mathbb R_+^l\times\mathbb Z_+^m$. The endowment mapping $e:I\to X$ is integrable. An individual commodity may have zero aggregate initial endowment.

\item There is a single firm owned by consumers.\footnote{The single-firm formulation is only to keep the notation simple.} Ownership shares are given by a measurable mapping $\theta:I\to\mathbb R_+$ with $\int_I\theta\,d\lambda=1$. The production set $Y\subseteq\mathbb R^{l+m}$ is closed and convex and satisfies: (i) $0\in Y$; (ii) $Y\cap(-Y)=\{0\}$ (irreversibility); (iii) $Y-\mathbb R_+^{l+m}\subseteq Y$ (free disposal); and (iv) the resource condition
\[
\left(Y+\int_I e\,d\lambda\right)\cap\mathbb R_{++}^{l+m}\neq\varnothing.
\]

\item An allocation is an $\mathcal I$-measurable, integrable function $f:I\to X$, assigning to each agent a bundle of divisible and indivisible goods.

\item The type space $T$ is complete, separable, and metrizable, with its Borel $\sigma$-algebra. \revision{The social type function $\alpha:I\to T$ is measurable. Let $\mathcal G\subseteq\mathcal I$ denote the $\sigma$-algebra generated by $\alpha$.} Agents evaluate the allocation through $\mathbb E(f\mid\alpha)$. The space of externalities is
\[
\revision{\Gamma=L_1((I,\mathcal G,\lambda);\mathbb R_+^{l+m}),}
\]
with the relative weak topology. Its values need not have integer coordinates: conditional averages of integer consumption can be fractional.

\item For each agent $i$, the conditional preference $\succ_i(\gamma)$ compares bundles at a given externality $\gamma$. We write $x\succ_i(\gamma)y$ when $x$ is strictly preferred to $y$. For every $\gamma$, this relation is irreflexive, transitive, and strictly monotone: $x\geq y$ and $x\neq y$ imply $x\succ_i(\gamma)y$. Its strict-preference graph is open in consumption and the externality; equivalently,
\[
\{(x,y,\gamma)\in X\times X\times\Gamma:
x\not\succ_i(\gamma)y\}
\]
is closed. No completeness assumption is imposed.

\item  {Let $\wp$ be the space of irreflexive, transitive preferences on $X$ with open strict-preference graphs, endowed with closed convergence of the complements of these graphs; see \citet[p.~96]{hi74}.} For each $\gamma$, the mapping $\Phi_\gamma:i\mapsto\succ_i(\gamma)$ is measurable into $\wp$, and
\[
\{(i,x,y):x\succ_i(\gamma)y\}
\in\mathcal I\otimes\mathcal B(X)\otimes\mathcal B(X).
\]

\item The conditional preference $\succ_i(\gamma)$ satisfies the compensability condition: for every $x\in X$
$$(c,0^m)\succ_i(\gamma) x \qquad \mbox{for some} \qquad c\in\mathbb R_+^l.
$$
 {Let $\wp_+\subseteq\wp$ denote the subset of strictly monotone preferences satisfying this compensability condition.}

\item  {Write $\mu_\gamma=\lambda\circ\Phi_\gamma^{-1}$. For every norm-bounded sequence $\{\gamma_n\}_{n\geq1}\subset\Gamma$, these distributions are uniformly tight: for each $\epsilon>0$, there is a compact set $K_\epsilon\subset\wp_+$ such that $\mu_{\gamma_n}(K_\epsilon)\geq1-\epsilon$ for every $n$.}

\item For every nonnegative, nonzero price vector $p$ at which $\pi(p):=\sup_{y\in Y}p\cdot y$ is finite, define the income distribution by
\[
\lambda_d^p(B)
=\lambda\{i\in I:p\cdot e(i)+\theta(i)\pi(p)\in B\},
\qquad B\in\mathcal B(\mathbb R).
\]
When $m\geq1$, this distribution is dispersed at every such price with $\sum_{h=1}^{l}p^h>0$ in the sense that $\lambda_d^p(\{b\})=0$ for all $b\in\mathbb R$.
\end{enumerate}

A generalized large economy is $(\mathcal E,\alpha,Y)$, where $\mathcal E(i)=(\succ_i,e(i),\theta(i))$. It extends the exchange model by allowing incomplete preferences, indivisible consumption, and production.

\begin{rmk}\label{rmk-gene cond}
\begin{itemize}
\item The production assumptions are standard; see \cite{hi70,hi74}. The firm can remain inactive, and free disposal allows unwanted goods to be discarded. Irreversibility rules out reversing a nonzero production plan at no cost. The aggregate resource condition requires that some feasible production plan make every commodity available in strictly positive aggregate amount. It permits zero initial supplies of goods that the firm can produce, as in the example below.

\item  Condition (8) requires that any bundle can be improved upon by sufficient consumption of divisible goods alone. It has been adopted in \cite{MC77} and \cite{HKH89}. 
Together with uniform tightness, it rules out a limiting price at which all divisible goods are free. When $m=0$, condition (8) follows from strict monotonicity. Taking $Y=-\mathbb R_+^{l+m}$ also admits pure exchange economies with indivisible goods, since dispersion is required only at prices that value a divisible good. 

\item With indivisible goods, demand can change discontinuously when income reaches the cost of a discrete bundle. If a positive mass of households has exactly that income, the discontinuity can persist in aggregate demand. The dispersed-income condition excludes such mass points.\footnote{For discussions of dispersion and nonconvex consumption, see \cite{KY81} and \cite{Suzuki(1995)}.} Because income includes distributed profits, the condition concerns total income rather than the value of initial endowments alone. No income-dispersion assumption is imposed when $m=0$.

\item Condition (10) can be weakened. For every nonnegative, nonzero $p$ with $\sum_{h=1}^{l}p^h>0$ and $\pi(p)<\infty$, it is enough to require
\begin{equation}\label{eq:critical-income}
\lambda\{i:p\cdot e(i)+\theta(i)\pi(p)=p\cdot(0^l,z)>0\}=0
\quad\text{for every }z\in\mathbb Z_+^m.
\end{equation}
Only positive incomes equal to the expenditure on a purely indivisible bundle need to have zero mass. At other positive incomes, a bundle on the budget boundary must contain positive expenditure on a divisible good, which can be reduced slightly when proving optimality. At strictly positive prices, a zero-income household has only the zero bundle available. Thus, \eqref{eq:critical-income} allows income atoms away from the critical expenditures, including an atom at zero. In the proof of Theorem~\ref{lct-ex2}, we indeed only use dispersion  in this weaker form. 
When $m=0$, it is vacuous.
\end{itemize}
\end{rmk}

We next define competitive equilibrium.

\begin{defn}
An allocation $f$, a nonnegative, nonzero price vector $p$, and a production plan $y\in Y$ form an equilibrium of $(\mathcal E,\alpha,Y)$ if the following conditions hold.
\begin{enumerate}
\item Individual optimality: for almost every $i$, $f(i)$ is maximal under $\succ_i(\mathbb E(f\mid\alpha))$ in
\[
B_i(p,y)=\{x\in X:p\cdot x\leq p\cdot e(i)+\theta(i)p\cdot y\}.
\]
That is, $f(i)\in B_i(p,y)$ and there is no $z\in B_i(p,y)$ with $z\succ_i(\mathbb E(f\mid\alpha))f(i)$.
\item Profit maximization: $p\cdot y=\sup_{y'\in Y}p\cdot y'$.
\item Market clearing: $\int_I f\,d\lambda=\int_I e\,d\lambda+y$.
\end{enumerate}
\end{defn}

The following characterization concerns the class of generalized economies.

\begin{thm}\label{lct-ex2}
\revision{Fix an agent space $(I,\mathcal I,\lambda)$ and a social type function $\alpha:I\to T$. The following statements are equivalent.}
\begin{enumerate}
\item[i.] \revision{The social type function  satisfies social anonymity.}
\item[ii.] Every generalized large economy admits an equilibrium.
\end{enumerate}
\end{thm}

\begin{rmk}\label{rmk-transformed}
 The externality can also be the conditional expectation of transformed consumption.  For a  concrete extension, consider  a continuous function $\tau:X\to\mathbb R^k$ satisfying
\[
\|\tau(x)\|\leq a+b\|x\|_1
\quad\text{for some }a,b\geq0,
\]
and let $\zeta=\mathbb E(\tau\circ f\mid\alpha)$ be the externality for each allocation $f$. Equip $L_1((I,\mathcal G,\lambda);\mathbb R^k)$ with its weak topology. Retain the assumptions of Section~\ref{glc}, with
conditions~(6)--(9) formulated on this externality space. Then social anonymity remains sufficient for equilibrium existence. \footnote{Such formulations also appear in large games; see, for example, \cite{Rauh2003}, \cite{YZ2005} and \cite{HS2018}.}

\end{rmk}

\subsection{An example}\label{subsec:general-example}

We now illustrate Theorem~\ref{lct-ex2} through a production economy in which households purchase cars. Cars provide practical benefits and social distinction. The latter depends on how common car ownership is among households of similar or lower wealth. The economy combines production, indivisible consumption, and social comparisons.

As in Section~\ref{sec-example}, let $I=[0,1]^2$ be equipped with the completion of product Lebesgue measure $\lambda$. A household is indexed by $(v,w)$, \revision{where $v$ is its initial wealth rank and $w$ distinguishes households with the same wealth rank.} \revision{Set $\alpha(v,w)=v$.} Social information records wealth rank and aggregates consumption across households at that rank. \revision{The independent, atomless household coordinate $w$ supplies individual variation beyond the social type, so social anonymity holds.}

There are two goods, with consumption set $X=\mathbb R_+\times\mathbb Z_+$. Good~1 is divisible consumption and good~2 is cars. Household endowments and ownership shares in the firm are $e(v,w)=(2+v,0)$ and $\theta(v,w)=v+\frac12$. Households with a higher wealth rank hold more of the divisible good and a larger ownership share. Hence
\[
\int_I e\,d\lambda=\left(\frac52,0\right),
\qquad
\int_I\theta\,d\lambda=1.
\]
The firm produces cars from the divisible good. Its production set is
\[
Y=\left\{(y^1,y^2)\in\mathbb R^2:
y^1\leq0,\quad y^1+y^2\leq0,\quad y^2\leq\frac12
\right\}.
\]
Producing one car requires at least one unit of the divisible good, aggregate production capacity is $1/2$, and free disposal is allowed. All cars available for consumption are produced by the firm. The production plan $(-\frac12,\frac12)$ leaves strictly positive aggregate resources of both goods.

A household compares car ownership with that of households whose wealth rank is no greater than its own. For $\gamma=(\gamma^1,\gamma^2)\in\Gamma$, define
\[
r_v(\gamma)=
\begin{cases}
\displaystyle\frac1v\int_0^v\int_0^1
\gamma^2(t,j)\,dj\,dt,&v>0,\\[1em]
\displaystyle\frac12,&v=0,
\end{cases}
\qquad
\eta_v(\gamma)=\min\{1,r_v(\gamma)\}.
\]
When $\gamma=\mathbb E(f\mid\alpha)$, $r_v(\gamma)$ is the average number of cars held by households in this comparison group. The value assigned at $v=0$ does not matter, because that group has measure zero.

Household $(v,w)$ has utility
\[
u_{(v,w)}(x,\gamma)
=x^1+\left(\frac32-\eta_v(\gamma)\right)
\frac{\log(1+x^2)}{\log 2}.
\]
The coefficient on car consumption can be written as
\[
\frac32-\eta_v(\gamma)
=\frac12+\bigl(1-\eta_v(\gamma)\bigr).
\]
The first term represents the private benefits of cars; the second represents a status premium. As car ownership becomes more widespread among households of similar or lower wealth, the distinction associated with ownership declines. The status premium vanishes when average ownership in the comparison group reaches one car per household. Cars continue to provide private benefits, and preferences remain strictly monotone. Additional cars yield diminishing benefits, while the integer constraint makes individual consumption choices nonconvex.

\begin{claim}\label{claim:car-equilibrium}
There exists an equilibrium with price $p^*=(1,1)$, production plan $y^*=(-\frac12,\frac12)$, and allocation
\[
f^*(v,w)=
\begin{cases}
(1+v,1),&0\leq w\leq\frac12,\\
(2+v,0),&\frac12<w\leq1.
\end{cases}
\]
\end{claim}

Half the households in each wealth group purchase one car, while the other half retain its cost in divisible consumption. Consequently,
\[
\mathbb E(f^*\mid\alpha)(v,w)
=\left(\frac32+v,\frac12\right),
\qquad
\eta_v\bigl(\mathbb E(f^*\mid\alpha)\bigr)=\frac12.
\]
At $p^*=(1,1)$, the firm earns zero profit and a household of wealth rank $v$ has income $2+v$. Purchasing $q$ cars and spending the remaining income on the divisible good gives utility
\[
2+v-q+\frac{\log(1+q)}{\log 2}.
\]
The choices $q=0$ and $q=1$ give the same utility, and every feasible integer $q\geq2$ gives strictly less. Both bundles assigned by $f^*$ are therefore optimal. The production plan $y^*$ maximizes profit, and markets clear:
\[
\int_I f^*\,d\lambda
=\left(2,\frac12\right)
=\left(\frac52,0\right)+\left(-\frac12,\frac12\right)
=\int_I e\,d\lambda+y^*.
\]

\revision{The example makes the role of social anonymity explicit.} Households with the same wealth rank can make different indivisible choices while generating a common comparison benchmark. \revision{Household identity permits these choices to vary among households with the same social type,} and the projection $\alpha(v,w)=v$ determines which information enters social comparisons. The resulting allocation supports both individual optimization and the social benchmarks on which those choices depend. We verify the model assumptions and prove the claim in  {Supplemental Appendix~\ref{app:car-proof}}.

\section{Concluding remarks}\label{sec-conclusion}

This paper develops a framework for competitive equilibrium in a large society in which the market allocation creates an endogenous social environment. The  modeling idea is to distinguish an agent's name from her social type. Names identify individuals, whereas social types determine the level at which individual choices are observed and aggregated into peer behavior, consumption benchmarks, and other externalities. The allocation determines the social environment, the social environment determines individual demand, and individual demand determines the allocation.

The central message is that, besides individual negligibility, the organization of the social environment matters for equilibrium existence. When social information is sufficiently rich to identify individual consumption, an equilibrium may fail to exist. When it leaves room for different individual choices while preserving the relevant social aggregates, conditional aggregation can reconcile nonconvex individual demand with market clearing. \revision{Social anonymity formalizes this distinction: social types never exhaust individual information on any positive-measure part of the population.} The rectangle example displays the distinction directly. Exact locations and  {educational attainment} generate comparisons that are incompatible with individual specialization; recording locations at the community level restores equilibrium.

\revision{Theorem~\ref{lct-ex1} identifies social anonymity as the condition under which every exchange economy admits an equilibrium.} The condition is necessary and sufficient for every fixed number of goods of at least two. Its necessity is established by a direct construction that keeps the commodity space fixed. Theorem~\ref{lct-ex2} extends the characterization to the class of generalized economies with incomplete preferences, production, and indivisible goods. This framework accommodates both preference nonconvexity and the intrinsic nonconvexity of individual consumption sets. It also permits zero initial supplies of goods that can be produced.

\revision{Taken together, the results suggest that a large economy with externalities should specify both the agent space and the social type function.} The former describes individual negligibility, while the latter determines which distinctions remain available when aggregating choices. This separates two questions: how the social environment is generated by equilibrium, and how the categories through which it is observed are specified as market primitives.

The applications illustrate this distinction in economies with peer effects, social comparison, and risk sharing with local social benchmarks. In each case, prices coordinate individual choices while the allocation reproduces the social objects entering preferences. The present paper abstracts from asymmetric and private information. A natural extension would introduce private information while retaining the social-type structure through which externalities are organized. We leave this extension for future work.

\appendix
\section{Appendix}\label{sec-appendix}

\subsection{Proof of Theorem~\ref{lct-ex1}}\label{app:proof-theorem1}

The sufficiency argument approximates prices while retaining the original endowments and full budget sets. Conditional convexification gives a fixed point at each approximation. Uniform tightness then rules out zero limiting prices, after which conditional selection yields an equilibrium allocation. For necessity, we place the rectangular economy of Section~\ref{sec-example} on a positive-measure set where social-type information is as rich as the full agent information.

\begin{proof}[Proof of sufficiency]
\revision{Suppose that $\alpha$ satisfies social anonymity, so $\mathcal I$ is nowhere equivalent to $\mathcal G=\sigma(\alpha)$.} Write $\bar e=\int_I e\,d\lambda, \boldsymbol{1}_l=(1,\ldots,1)\in\mathbb R^l$ and let
\[
\Delta=\left\{p\in\mathbb R_+^l:\sum_{j=1}^l p^j=1\right\},
\qquad p^0=l^{-1}\boldsymbol 1_l,
\qquad \Delta_\epsilon=(1-\epsilon)\Delta+\epsilon p^0,
\quad 0<\epsilon<1.
\]

\medskip
\noindent\textit{Step 1. Auxiliary equilibria and bounded aggregate consumption.}
For $p\gg0$, retain the full budget set $B(i,p)=\{x\in\mathbb R_+^l:p\cdot x\le p\cdot e(i)\}$ and demand $D(i,p,\gamma)$ from Section~\ref{subsec-setup}. For each $i$, the budget correspondence $p\mapsto B(i,p)$
is continuous and has nonempty compact values on
$\mathbb R_{++}^l$. Berge's maximum theorem and the measurable maximum theorem
imply that $D(i,p,\gamma)$ is nonempty and compact,
$i\mapsto D(i,p,\gamma)$ is measurable for each $(p,\gamma)$,
and $(p,\gamma)\mapsto D(i,p,\gamma)$ is upper hemicontinuous
for each $i$.
For $p\in\Delta_\epsilon$, every budget-feasible bundle satisfies
\[
\|x\|_1\le b_\epsilon(i):=\frac l\epsilon\|e(i)\|_1.
\]
Define
\[
\Gamma_\epsilon=
\left\{\gamma\in\Gamma:0\le\gamma^j\le\mathbb E(b_\epsilon\mid\alpha)
\text{ a.e.},\ j=1,\ldots,l\right\}.
\]
The common integrable bound makes this set uniformly integrable. It is also convex and weakly closed, so the Dunford--Pettis theorem gives weak compactness. Since $\mathcal G$ is countably generated and
$\Gamma_\epsilon$ is weakly compact, the relative weak
topology on $\Gamma_\epsilon$ is metrizable. By \citet[Theorems~2--4]{HS2018}, the correspondence
\[
\phi_\epsilon(p,\gamma)
=\{\mathbb E(f\mid\alpha):f(i)\in D(i,p,\gamma)\text{ a.e.}\}
\]
has nonempty, convex, weakly compact values in $\Gamma_\epsilon$ and is upper hemicontinuous. For $\gamma\in \Gamma_\epsilon$, let
\[
\psi_\epsilon(\gamma)
=\argmax_{q\in\Delta_\epsilon}
q\cdot\left(\int_I\gamma\,d\lambda-\bar e\right).
\]
Since $\Delta_\epsilon$ is nonempty, compact and convex
and the objective is linear in $q$, $\psi_\epsilon$ has
nonempty, convex, compact values. Moreover,
$\gamma\mapsto\int_I\gamma\,d\lambda$ is weakly continuous.
Berge's maximum theorem therefore gives upper hemicontinuity
of $\psi_\epsilon$ under the weak $L^1$ topology on
$\Gamma_\epsilon$. The Fan--Glicksberg theorem gives a fixed point $(p_\epsilon,\gamma_\epsilon)$ of $\psi_\epsilon\times\phi_\epsilon$ and an allocation $f_\epsilon$ such that
$\gamma_\epsilon=\mathbb E(f_\epsilon\mid\alpha)$ and $f_\epsilon(i)\in D(i,p_\epsilon,\gamma_\epsilon)$ almost everywhere. Strict monotonicity implies budget exhaustion. Thus, with $z_\epsilon=\int_I \gamma_\epsilon\,d\lambda-\bar e$,
\begin{equation}\label{eq:exchange-auxiliary-prices}
p_\epsilon\cdot z_\epsilon=0,
\qquad q\cdot z_\epsilon\le0\quad(q\in\Delta_\epsilon).
\end{equation}
In particular, $p^0\in\Delta_\epsilon$ gives
\begin{equation}\label{eq:aggregate-bound}
\int_I\|f_\epsilon(i)\|_1\,d\lambda
\le\int_I\|e(i)\|_1\,d\lambda.
\end{equation}
Since $\gamma_\epsilon$ is nonnegative and has the same integral as $f_\epsilon$, it satisfies the same $L_1$ bound.

\medskip
\noindent\textit{Step 2. Strict positivity of the limiting price.}
Choose $\epsilon_n\downarrow0$ and a subsequence such that $p_n\to p\in\Delta$, writing $f_n$ and $\gamma_n$ for the corresponding allocations and externalities. Let $w_n(i)=p_n\cdot e(i)$ and $w(i)=p\cdot e(i)$. Since $\bar e\gg0$, the set $A=\{i:w(i)>0\}$ has measure $\delta>0$.

Suppose that $p^j=0$. Using \eqref{eq:aggregate-bound}, Markov's inequality,
and uniform tightness, choose $R<\infty$ and a compact set $K\subset\wp_+$ such that
\[
\lambda\{i:\|f_n(i)\|_1>R\}<\delta/4,
\qquad
\lambda\{i:u_i(\cdot,\gamma_n)\notin K\}<\delta/4
\]
for every $n$. Consequently,
\[
S_n=A\cap\{i:\|f_n(i)\|_1\le R\}
\cap\{i:u_i(\cdot,\gamma_n)\in K\}
\]
has measure at least $\delta/2$. Since $\lambda(\limsup_n S_n)\ge\limsup_n\lambda(S_n)$, a positive-measure set of agents belongs to infinitely many $S_n$. Discarding the union of the null sets where individual optimality fails, choose one such agent $i$. Compactness of $\{x\geq0:\|x\|_1\leq R\}$ and of $K$, together with budget exhaustion, gives a subsequence along which
\[
f_n(i)\to x,\qquad u_i(\cdot,\gamma_n)\to v\in K,
\qquad p\cdot x=w(i)>0,
\]
where utility convergence is uniform on compact consumption sets. Let $q^j$ denote the $j$th coordinate unit vector. Strict monotonicity gives $v(x+q^j)>v(x)$. By continuity, choose $\chi\in(0,1)$ sufficiently small that
\[
v((1-\chi)x+q^j)>v(x),
\qquad
p\cdot((1-\chi)x+q^j)=(1-\chi)w(i)<w(i).
\]
Price and utility convergence imply that $(1-\chi)x+q^j$
is affordable and strictly preferred to $f_n(i)$ for all
sufficiently large indices along the subsequence,
a contradiction. Therefore $p\gg0$.

\medskip
\noindent\textit{Step 3. The limiting allocation.}
By Step $2$, every coordinate of $p_n$ is bounded below by some $\beta>0$ for all sufficiently large $n$. Budget feasibility then gives the common integrable bound $\|f_n(i)\|_1\le\beta^{-1}\|e(i)\|_1$. The externalities lie in a common weakly compact order interval.
Hence, after passing to a subsequence,
$
\gamma_n=\mathbb E(f_n\mid\alpha)\rightharpoonup\gamma$ for some $\gamma\in\Gamma.$ By \citet[Lemma~3]{HS2018}, there is an allocation $f$ satisfying
$
f(i)\in\ls_n f_n(i)$ a.e.,
$\mathbb E(f\mid\alpha)=\gamma,
$
where $\ls_n x_n$ denotes the set of limits of convergent
subsequences of the vector sequence $(x_n)$. 

For almost every $i$, take a subsequence with $f_n(i)\to f(i)$. Since $p_n\to p\gg0$ and $\gamma_n\rightharpoonup\gamma$,
upper hemicontinuity of $D(i,\cdot,\cdot)$ gives
$f(i)\in D(i,p,\gamma)$.
Since $\gamma_n\rightharpoonup\mathbb E(f\mid\alpha)$, $z_n=\int_I\gamma_n\,d\lambda-\bar e
\to
\int_I f\,d\lambda-\bar e=:z.$  For any $q\in\Delta$, apply \eqref{eq:exchange-auxiliary-prices} to $(1-\epsilon_n)q+\epsilon_n p^0$ and pass to the limit to obtain $q\cdot z\le0$ for every $q\in\Delta$ and $p\cdot z=0$. Hence $z\le0$; since $p\gg0$, $z=0$. Thus markets clear, and $(f,p)$ is an equilibrium.
\end{proof}

\paragraph{The ordinal condition in Remark~\ref{rmk-cover}.}
For completeness, the same proof works under \eqref{eq:ordinal-stability} in place of uniform tightness. Only Step~2 changes. Choose a positive-measure set $A$ on which $w(i)>0$, and use the aggregate bound to choose $R$ so that $\|f_n(i)\|_1\le R$ outside a set of sufficiently small measure. Apply \eqref{eq:ordinal-stability} with an exceptional set smaller than $\lambda(A)/4$. 
There is an agent $i\in A$ and an infinite subsequence along which
$\|f_n(i)\|_1\le R$ and the inequalities in (1) hold for that agent.
For this agent, budget exhaustion and $w_n(i)\to w(i)>0$ imply that
\[
b_n=(1-\eta)f_n(i)+q^j,
\qquad
p_n\cdot b_n=w_n(i)-\eta w_n(i)+p_n^j<w_n(i)
\]
for all sufficiently large indices whenever $p^j=0$. Along the selected subsequence, the ordinal condition makes $b_n$ strictly preferred to $f_n(i)$, a contradiction. The remainder of the proof is unchanged.

We next prove necessity. The following bound supplies the contradiction in the rectangular example and in its embedding below.

\begin{lma}[Rectangle bound]\label{lem:rectangle}
Let $q:[0,1]^2\to[-1,1]$ be measurable and define
\[
H(v,w)=c+\int_0^v\int_0^w q(t,j)\,dj\,dt,
\qquad c\in\mathbb R.
\]
Suppose that $q=-1$ almost everywhere on $\{H>0\}$ and $q=1$ almost everywhere on $\{H<0\}$. Then $|H(v,w)|\le |c|$ everywhere. In particular, if $c=0$, then $H\equiv0$ and $q=0$ almost everywhere.
\end{lma}

\begin{proof}
The integral representation makes $H$ Lipschitz continuous, with $H_{vw}=q$ almost everywhere and $H(v,0)=H(0,w)=c$. We first handle $c=0$ by an integral identity. 
For $c>0$, we construct a function $\phi$ with $|\phi(s)|\le c$ and then prove that $H(v,w)=\phi(vw)$.
The case $c<0$ follows by replacing $(H,q,c)$ with $(-H,-q,-c)$.

\medskip
\noindent\textit{Step 1. Zero boundary values.}
Suppose that $c=0$. The sign restriction determines $q$ only where $H\ne0$. To obtain an identity valid almost everywhere on the whole square, we first show that
 \begin{equation}\label{eq:H-zero-set}
H_v=0\quad\text{a.e. on }\{H=0\}.
\end{equation}

To see this, fix $w$ and write $h(v)=H(v,w)$. Since $h$ is Lipschitz,
$h'$ exists almost everywhere. At a non-isolated zero $v$ where
$h'$ exists, choose distinct zeros $v_n\to v$. Then
\[
h'(v)=\lim_{n\to\infty}
\frac{h(v_n)-h(v)}{v_n-v}=0.
\]
The isolated zeros are countable, and the points where $h'$ does
not exist form a null set. Thus $h'=0$ almost everywhere on
$\{h=0\}$. At any zero where $h'(v)=0$, we have
$h(v+\varepsilon)=o(|\varepsilon|)$, which also gives
$(|h|)'(v)=0$. Away from the zero set, the ordinary chain rule
applies wherever $h'$ exists. Consequently,
$(|h|)'=\operatorname{sgn}(h)h'$ almost everywhere, where
$\operatorname{sgn}(0)=0$. These conclusions hold for every fixed $w$.
Let $N$ consist of points where $H_v$ does not exist,
or where $H=0$ and $H_v$ exists but is nonzero.
The preceding argument shows that for each fixed $w$, the section $N_w=\{v:(v,w)\in N\}$ has one-dimensional measure zero.
Fubini's theorem therefore gives
\[
\iint_{[0,1]^2}\mathbf{1}_N(v,w)\,dv\,dw
=
\int_0^1
\underbrace{\left(\int_0^1\mathbf{1}_N(v,w)\,dv\right)}_{=\,0}
\,dw
=0.
\]
Thus $N$ has two-dimensional measure zero. Outside $N$, $H=0$ implies $H_v=0$, proving
\eqref{eq:H-zero-set}.

Consequently,
\begin{equation}\label{eq:absH-derivative}
\partial_v|H|=\operatorname{sgn}(H)H_v
\quad\text{a.e. on }[0,1]^2.
\end{equation}

For almost every $v$, we take the absolutely continuous
representative $H_v(v,w)=\int_0^w q(v,j)\,dj$. Then
\begin{equation}\label{eq:Hv-square-derivative}
\partial_w\left(\frac12H_v^2\right)=H_vq
\quad\text{a.e. on }[0,1]^2.
\end{equation}
On $\{H\ne0\}$, the sign restriction gives
$q=-\operatorname{sgn}(H)$ almost everywhere; on $\{H=0\}$,
\eqref{eq:H-zero-set} gives $H_v=0$ almost everywhere.
Combining \eqref{eq:absH-derivative} and
\eqref{eq:Hv-square-derivative}, we obtain
\begin{equation}\label{eq:H-energy-identity}
\partial_w\left(\frac12H_v^2\right)+\partial_v|H|
=H_v\bigl[q+\operatorname{sgn}(H)\bigr]=0
\quad\text{a.e. on }[0,1]^2.
\end{equation}

Integrating \eqref{eq:H-energy-identity} over
$[0,v]\times[0,w]$, using $H_v(t,0)=H(0,j)=0$, gives
\[
\frac12\int_0^v H_v(t,w)^2\,dt
+\int_0^w|H(v,j)|\,dj=0
\]
for every pair of endpoints. Both integrals must vanish, so $H=0$ almost everywhere. Continuity gives $H\equiv0$, and hence $q=H_{vw}=0$ almost everywhere.

\medskip
\noindent\textit{Step 2. Constructing and bounding the profile for $c>0$.}
We begin by identifying a region on which $H$ is already determined. Since $q\ge-1$, we have $H(v,w)\ge c-vw$. Hence $H>0$ whenever $vw<c$, and $q=-1$ almost everywhere in this region. Every lower rectangle ending at a point with $vw\le c$ lies in this region up to a null boundary. Therefore
\[
H(v,w)=c-vw\qquad\text{whenever }vw\le c.
\]
If $c\ge1$, this determines $H$ on the whole square and proves the bound. Assume henceforth that $0<c<1$.

The solution thus depends only on $vw$ up to the product level $c$. We construct a continuation of this one-variable profile, beginning with
\[
\phi(s)=c-s,\qquad 0\le s\le c.
\]
Its first zero is $c$, with $\phi'(c)=-1$. We extend $\phi$ by joining pieces with matching values and first derivatives.


Suppose a constructed piece ends at a zero $r>0$ with slope
$d\ne0$. Set $\sigma=\operatorname{sgn}(d)\in\{-1,1\}$.
To match this slope, the next piece must be positive immediately
to the right of $r$ if $d>0$, and negative if $d<0$.
Thus $\sigma$ specifies the sign with which the next piece
leaves zero. Since $H(v,w)=c-vw$ on the initial region, we seek a continuation
of the form $H^*(v,w)=\phi(vw)$. Writing $s=vw$, we have
$\partial_{vw}H^*=(s\phi'(s))'$. On a piece with sign $\sigma$,
the sign restriction therefore requires $(s\phi'(s))'=-\sigma$.
Integrating twice gives
\[
\phi(s)=A+B\log s-\sigma s.
\]
 Choose $A$ and $B$ to match the
value and slope at $r$,  which gives
\begin{equation}\label{eq:phi-matching}
A+B\log r-\sigma r=0,
\qquad \frac{B}{r}-\sigma=d.
\end{equation}
Solving \eqref{eq:phi-matching} gives
$
B=r(d+\sigma), A=\sigma r-r(d+\sigma)\log r.
$
Thus the next piece is uniquely determined:
\begin{equation}\label{eq:phi-piece}
\phi(s)
=r(d+\sigma)\log(s/r)-\sigma(s-r)
=\sigma\bigl[r(1+|d|)\log(s/r)-(s-r)\bigr],
\end{equation}
where $d=\sigma|d|$.
Equation \eqref{eq:phi-piece} applies while $\phi$ has sign
$\sigma$, so we use it until the next zero or $s=1$,
whichever comes first. For $0<c<1$, the initial piece
$\phi(s)=c-s$ gives $r=c$, $d=-1$, and $\sigma=-1$.
Substituting into \eqref{eq:phi-piece} yields
\begin{equation}\label{eq:phi-first-piece}
\phi(s)=s-c-2c\log(s/c).
\end{equation}
The expression in \eqref{eq:phi-first-piece}
has derivative $1-2c/s$.
It therefore decreases from zero at $s=c$ to the negative
minimum $c(1-2\log2)$ at $s=2c$, and then increases.
Since \eqref{eq:phi-first-piece} tends to $+\infty$ as $s\to\infty$,
it has a unique next zero $r_2>2c$.
We use this piece only on $[c,\min\{r_2,1\}]$.
To verify that the construction can be repeated, let
$a=r(1+|d|)>r$ and $F(s)=a\log(s/r)-(s-r).$ Then
$\phi(s)=\sigma F(s), F'(s)=\frac{a}{s}-1.$
Since $F(r)=0$, $F$ increases on $(r,a)$ and then decreases
to $-\infty$. Thus there is a unique $t>r$ such that
\begin{equation}\label{eq:next-zero-location}
F(t)=0,\qquad t>a=r(1+|d|).
\end{equation}
Moreover, $F>0$ on $(r,t)$. Hence $\phi$ has sign $\sigma$
on $(r,t)$, and
\[
\phi'(t)=\sigma\left(\frac{a}{t}-1\right)\ne0,
\qquad
\operatorname{sgn}\phi'(t)=-\sigma=-\operatorname{sgn}(d).
\]
If $t<1$, the same rule therefore defines the next piece.

The pieces match in value and slope at each joining point,
so the constructed function is $C^1$ after any finite number
of steps. On each open sign interval,
$\phi(s)=A+B\log s-\sigma s$, and hence
\[
\phi'(s)=\frac Bs-\sigma,
\qquad
\phi''(s)=-\frac B{s^2}.
\]
Therefore,
\begin{equation}\label{eq:phi-ode}
s\phi''(s)+\phi'(s)=-\sigma=-\operatorname{sgn}\phi(s).
\end{equation}
This equation holds inside each piece; $\phi''$ need not exist
at the joining points. To show that the construction reaches
$s=1$, we next bound $\phi$ and $\phi'$ and rule out
an accumulation of zeros.

Define
\[
E(s)=\frac{s}{2}\phi'(s)^2+|\phi(s)|.
\]
On each open interval where $\phi$ has a constant nonzero sign, \eqref{eq:phi-ode} gives
\begin{align*}
E'(s)
&=\frac12\phi'(s)^2+s\phi'(s)\phi''(s)
  +\operatorname{sgn}(\phi(s))\phi'(s)\\
&=\frac12\phi'(s)^2
  +\phi'(s)\bigl[s\phi''(s)+\operatorname{sgn}\phi(s)\bigr]
=-\frac12\phi'(s)^2.
\end{align*}
The identity also holds at each zero. Indeed, $E$ is continuous at a joining point $r$ because $\phi$ and $\phi'$ are continuous there. For sufficiently small $h\ne0$, the mean value theorem gives $\xi_h$ between $r$ and $r+h$ such that
\[
\frac{E(r+h)-E(r)}h=-\frac12\phi'(\xi_h)^2.
\]
Letting $h\to0$ gives $E'(r)=-\frac12\phi'(r)^2$. Hence $E$ is continuously differentiable on each finite constructed portion, including its joining points, with one-sided derivatives at endpoints.

Since $E$ is nonincreasing, we have $|\phi(s)|\le E(s)\le E(0)=c.$
To control the distance between successive zeros, observe also that
\[
(sE(s))'=E(s)+sE'(s)=|\phi(s)|\ge0.
\]
At the first zero $s=c$, $E(c)=c/2$. At any constructed zero $r\ge c$, write $\phi'(r)=d$. Then
\[
\frac{(rd)^2}{2}=rE(r)\ge cE(c)=\frac{c^2}{2},
\qquad\text{so}\qquad r|d|\ge c.
\]
By \eqref{eq:next-zero-location}, the next zero $t$, if reached, occurs after the turning point $r(1+|d|)$, and hence
$
t-r>r|d|\ge c.
$
Successive zeros are separated by more than $c$, so only finitely many can occur in $[0,1]$. Each explicit piece is finite at every finite positive argument. Thus the construction defines $\phi\in C^1([0,1])$ with
$|\phi(s)|\le c$ for every $s\in[0,1]$.
The function has only finitely many zeros, and
$\phi'(r)\ne0$ at each zero $r$.

\medskip
\noindent\textit{Step 3. Showing that $H$ follows the constructed profile.}
Set $H^*(v,w)=\phi(vw)$ and $q^*(v,w)=-\operatorname{sgn}\phi(vw)$ for $v,w\in[0,1]$. We now prove that $H=H^*$ on $[0,1]^2$.
Step 2 already gives this equality whenever $vw\le c$.

 On the initial piece, $\phi''=0$. On each later piece,
\eqref{eq:phi-piece} gives
\[
\phi''(s)=-\frac{r(d+\sigma)}{s^2},
\]
which is bounded since $s\ge r>0$.
Since there are finitely many pieces and $\phi'$ is continuous
at their joining points, $\phi'$ is Lipschitz and hence
absolutely continuous. By \eqref{eq:phi-ode},
\[
H^*_{vw}(v,w)=\phi'(vw)+vw\phi''(vw)=q^*(v,w)
\quad\text{a.e.}
\]
Integrating $H^*_{vw}=q^*$ over $[0,v]\times[0,w]$ and using $H^*(v,0)=H^*(0,w)=c$ yields
\[
H^*(v,w)=c+\int_0^v\int_0^w q^*(t,j)\,dj\,dt.
\]
Thus $H$ and $H^*$ both equal $c$ when $v=0$ or $w=0$ and have double-integral representations with integrands
$q$ and $q^*$, respectively. Step~2 established their equality whenever $vw\le c$.

Let
\[
\rho=\sup\{a\in[0,1]:H(v,w)=\phi(vw)\text{ whenever }vw\le a  \text{ and } (v,w)\in [0,1]\times[0,1]\}.
\]
We have $\rho\ge c>0$. By the definition of $\rho$, equality holds whenever $vw<\rho$.
If $vw=\rho$, take $v_n=(1-1/n)v$. Then $v_nw<\rho$, and
continuity gives
$
H(v,w)=\lim_{n\to\infty}H(v_n,w)
=\lim_{n\to\infty}\phi(v_nw)=\phi(\rho).
$
Thus $H(v,w)=\phi(vw)$ whenever $vw\le \rho$. Suppose that $\rho<1$. On $\{vw\le \rho\}$, $H=H^*$, so the sign restrictions give
$q=q^*$ almost everywhere where $H^*\ne0$.
Since $\phi$ has finitely many zeros $r_k$, the zero set of
$H^*$ is a finite union of sets $\{vw=r_k\}$, each of
two-dimensional measure zero.
Hence $q=q^*$ almost everywhere on $\{vw\le \rho\}$.

For $s=vw>\rho$, let $D=[0,v]\times[0,w]\cap\{tj>\rho\}$.
Since $q=q^*$ almost everywhere on
$([0,v]\times[0,w])\setminus D$, the integral of $q-q^*$
over this set is zero. The vertical section of $D$ at $t$ is $(\rho/t,w]$ for $\rho/w<t\le v$
and is empty otherwise. Hence $|q-q^*|\le2$ gives
\begin{align}\label{eq:est_diff}
|H(v,w)-\phi(s)|
&\le\iint_D|q-q^*|\,dt\,dj
\le2\int_{\rho/w}^{v}\left(w-\frac \rho t\right)\,dt\notag\\
&=2\bigl[s-\rho-\rho\log(s/\rho)\bigr]
=2\int_\rho^s\frac{u-\rho}{u}\,du\notag\\
&\le\frac2\rho\int_\rho^s(u-\rho)\,du
=\frac{(s-\rho)^2}{\rho}.
\end{align}
This estimate depends only on $s$, so it holds uniformly over all points at the same product level.

If $\phi(\rho)\ne0$, continuity gives
\[
|\phi(s)|\ge\frac{|\phi(\rho)|}{2}>
\frac{(s-\rho)^2}{\rho}
\]
for $s>\rho$ sufficiently close to $\rho$.
If $\phi(\rho)=0$, then $\phi'(\rho)\ne0$ and
$\phi(s)/(s-\rho)\to\phi'(\rho)$ as $s\downarrow\rho$. Hence
\[
|\phi(s)|\ge\frac{|\phi'(\rho)|}{2}(s-\rho)>
\frac{(s-\rho)^2}{\rho}
\]
for $s>\rho$ sufficiently close to $\rho$.
In either case, choose $\delta\in(0,1-\rho)$ small enough
that \eqref{eq:est_diff} implies
\begin{equation}\label{eq:sign-comparison}
|H(v,w)-\phi(s)|
\le\frac{(s-\rho)^2}{\rho}<|\phi(s)|,
\qquad \rho<s=vw\le\rho+\delta.
\end{equation}
By \eqref{eq:sign-comparison}, if $\phi(s)>0$, then
\[
H(v,w)\ge\phi(s)-|H(v,w)-\phi(s)|
>\phi(s)-|\phi(s)|=0.
\]
Similarly, if $\phi(s)<0$, then
\[
H(v,w)\le\phi(s)+|H(v,w)-\phi(s)|
<\phi(s)+|\phi(s)|=0.
\]
Thus $H(v,w)$ and $\phi(vw)$ have the same nonzero sign
whenever $\rho<vw\le\rho+\delta$.

Their sign restrictions now force $q=q^*$ almost everywhere on $\{vw\le \rho+\delta\}$. Every lower rectangle ending in this region lies entirely within it. Integrating the equal integrands gives $H(v,w)=H^*(v,w)=\phi(vw)$ whenever $vw\le \rho+\delta$, contradicting the definition of $\rho$. Thus $\rho=1$ and $|H(v,w)|=|\phi(vw)|\le c$ throughout the square. Together with Step~1 and sign reversal for $c<0$, this proves the lemma.
\end{proof}

\begin{proof}[Proof of necessity in Theorem~\ref{lct-ex1}]
\revision{Fix $l\ge2$ and the given social type function $\alpha:I\to T$, and suppose that social anonymity fails. Write $\mathcal G=\sigma(\alpha)$. There is $D\in\cI$, with $\zeta=\lambda(D)>0$, such that every measurable subset of $D$ agrees modulo null sets with $D\cap G$ for some $G\in\cG$. We embed the rectangular example on $D$, keeping the agent space, the social type function, and the number of goods fixed.}

\medskip
\noindent\textit{Step 1. Representing the subeconomy on the unit square.}
 Let $\lambda_D=\lambda|_D/\zeta$.
The probability space $(D,\cG^D,\lambda_D)$ is atomless:
any positive-measure set can be split in $\cI^D$, and
the resulting subsets have representatives in $\cG^D$
modulo null sets. Since $\cG^D$ is countably generated,
\citet[Lemma~6]{HSS2017} gives a measure-preserving map from $(D,\cG^D,\lambda_D)$ to the Lebesgue unit interval that generates $\cG^D$
modulo null sets. Composing it with a measure-preserving
Borel isomorphism from $[0,1]$ to $[0,1]^2$
\citep[Theorem~3.4.23]{sr98}, we obtain a
$\cG^D$-measurable map $\psi:D\to[0,1]^2$ satisfying $\lambda_D\circ\psi^{-1}=\operatorname{Leb}_{[0,1]^2}, \sigma(\psi)=\cG^D$ modulo $\lambda_D$-null sets. Since $[0,1]^2$ is Polish,
\citet[Proposition~3.3.4]{sr98} gives a $\cG$-measurable
extension $\varphi:I\to[0,1]^2$ with $\varphi|_D=\psi$.
\revision{Since $\varphi$ is $\sigma(\alpha)$-measurable, there is a Borel map $\vartheta:T\to[0,1]^2$ such that $\varphi=\vartheta\circ\alpha$.}
For $z=(v,w)$, define
\[
B_z=\{i\in I:\varphi^1(i)\le v,\ \varphi^2(i)\le w\}
\in\cG.
\]
\revision{Thus, $B_z=\alpha^{-1}(\vartheta^{-1}([0,v]\times[0,w]))$: the comparison groups are determined by the original social types. The map $\varphi$ supplies auxiliary coordinates and does not replace $\alpha$.}

\medskip
\noindent\textit{Step 2. Constructing and verifying the economy.}
For $x\in\mathbb R_+^l$ and any $\mathcal I$-measurable
map $h:I\to\mathbb R_+^l$, define
\begin{equation}\label{eq:composite-quantities}
\widehat x
=\left(x^1,\frac1{l-1}\sum_{j=2}^l x^j\right),
\qquad
\widehat h(i)
=\left(h^1(i),\frac1{l-1}\sum_{j=2}^l h^j(i)\right).
\end{equation}

Set $e(i)=1_D(i)\boldsymbol{1}_l$. Recall that
$\Gamma=L_1((I,\mathcal G,\lambda);\mathbb R_+^l)$,
endowed with the relative weak topology.  For
$z\in[0,1]^2$ and $\gamma\in\Gamma$, define
\[
A_k(z;\gamma)
=\frac1\zeta\int_{B_z}\widehat\gamma^k\,d\lambda,
\qquad k=1,2,
\]
 where $\widehat\gamma$ is defined by
\eqref{eq:composite-quantities}.
For $(x,\gamma)\in\mathbb R_+^l\times\Gamma$, let
\[
u_i(x,\gamma)
=\sum_{k=1}^2
\left(
\frac{\widehat x^k}{1+A_k(\varphi(i);\gamma)}
\right)^{3/2}.
\]
The aggregate endowment is $\zeta\boldsymbol{1}_l\gg0$.
For each fixed $z$, $A_k(z;\cdot)$ is a bounded linear
functional on $L^1$. Hence $u_i$ is jointly continuous
in $(x,\gamma)$. Since $(1+A_k(\varphi(i);\gamma))^{-3/2}>0$
for $k=1,2$ and increasing any commodity strictly increases
one composite quantity, $u_i$ is strictly monotone
in every commodity. For fixed $\gamma$, the function
$(z,i)\mapsto\mathbf1_{B_z}(i)\widehat\gamma^k(i)$
is jointly measurable. Integrating with respect to $i$
therefore shows that $z\mapsto A_k(z;\gamma)$ is
Borel measurable. Since $\varphi$ is measurable,
$i\mapsto u_i(x,\gamma)$ is measurable for every
fixed $(x,\gamma)$.

If $\|\gamma_n\|_{L^1}\le M$ for all $n$, then
$
0\le A_k(z;\gamma_n)\le M/\zeta.
$
Thus all conditional utilities belong to 
\[
\mathcal K_b
:= \left\{
x\to
b_1(\widehat x^1)^{3/2}+b_2(\widehat x^2)^{3/2}:
(b_1,b_2)\in[b,1]^2
\right\},
\qquad \text{where} \qquad
b=(1+M/\zeta)^{-3/2}>0.
\]
Since $\mathcal K_b$ is the continuous image of $[b,1]^2$,
it is compact in the compact-open topology. Since all conditional utilities lie in the common compact set
$\mathcal K_b$ of strictly monotone utilities, uniform tightness follows.

\medskip
\noindent\textit{Step 3. Prices and consumption choices in equilibrium.}
Suppose that $(f,p)$ is an equilibrium. Strict monotonicity
requires $p\gg0$. Agents outside $D$ have zero income and
therefore consume zero. The prices of goods $2,\ldots,l$ must be equal. Otherwise,
replacing a positive amount of a more expensive good by a cheaper
substitute and spending the savings on good~1 would strictly
increase utility. The more expensive good would therefore have
zero demand, contrary to its positive aggregate supply.

Since $p^2=\cdots=p^l$, define the composite prices by 
$\widetilde p
=\bigl(p^1,(l-1)p^2\bigr).$
Then
\[
p\cdot x
=p^1x^1+p^2\sum_{j=2}^l x^j
=\widetilde p^1\widehat x^1
+\widetilde p^2\widehat x^2.
\]
Normalize prices so that
$\widetilde p^1+\widetilde p^2=1$. Every agent in $D$ then has income one. 
Fix the equilibrium externality
$\gamma=\mathbb E(f\mid\alpha)$.
For any $a=(a^1,a^2)\in\mathbb R_+^2$ with
$\widetilde p\cdot a\le1$, the original consumption bundle
$
\xi(a)=(a^1,a^2,\ldots,a^2)\in\mathbb R_+^l
$
satisfies $\widehat{\xi(a)}=a$ and
$p\cdot\xi(a)=\widetilde p\cdot a\le1$. 
Thus $\xi(a)$ is affordable for every agent in $D$.
Individual optimality in the original economy implies
$u_i(f(i),\gamma)\ge u_i(\xi(a),\gamma)$ for every such $a$ and almost every $i\in D$. Every affordable composite bundle $a$ has an affordable
realization $\xi(a)$ in the original economy, with
$\widehat{\xi(a)}=a$. Let $\widehat f$ be the composite allocation defined by
\eqref{eq:composite-quantities}. Since utility depends only on the
composite quantities, individual optimality implies that
$\widehat f(i)$ maximizes the induced utility over
$\{a\in\mathbb R_+^2:\widetilde p\cdot a\le1\}$.
Budget exhaustion gives $\widetilde p\cdot\widehat f(i)=1$.
Any bundle on the budget line with both coordinates positive
lies strictly between the two corners. Strict convexity of
utility in the composite quantities therefore makes at least
one corner strictly preferable. Consequently,
\[
\widehat f(i)\in
\left\{
(1/\widetilde p^1,0),
(0,1/\widetilde p^2)
\right\}
\quad\text{for almost every }i\in D.
\]

\medskip
\noindent\textit{Step 4. Applying the rectangle bound.}
Let
$
D_1=\{i\in D:\widehat f^1(i)>0\}.
$
By the defining property of $D$  and the generating property of
$\varphi|_D$, there is a Borel set $C\subseteq[0,1]^2$ such that
$D_1=D\cap\varphi^{-1}(C)$ modulo null sets. Define
\[
g(z)=
\begin{cases}
(1/\widetilde p^1,0),&z\in C,\\
(0,1/\widetilde p^2),&z\notin C.
\end{cases}
\]
Then $\widehat f=g\circ\varphi$ almost everywhere on $D$. For $z=(v,w)$ and $k=1,2$,
\begin{align}\label{eq:rectangle-aggregates}
&A_k(z;\mathbb E(f\mid\alpha))
=\frac1 \zeta \int_{B_z}
  \mathbb E(\widehat f^k\mid\alpha)\,d\lambda
=\frac1\zeta\int_{B_z}\widehat f^k\,d\lambda\notag\\
&=\frac1\zeta\int_{D\cap B_z}g^k(\varphi(i))\,d\lambda(i)
=\int_0^v\int_0^w g^k(t,j)\,dj\,dt
=:F_k(v,w).
\end{align}
 The first equality uses the definition of $A_k$ and the
linearity of $x\mapsto\widehat x$. The second follows from
$B_z\in\cG$ and the property of conditional expectation.
The third uses $f=0$ outside $D$ and $\widehat f=g\circ\varphi$
on $D$, up to null sets. The fourth follows from
$B_z=\varphi^{-1}([0,v]\times[0,w])$ and the uniform distribution
of $\varphi$ under $\lambda_D=\lambda|_D/\zeta$. 

Taking $z=(1,1)$ in \eqref{eq:rectangle-aggregates} gives
\begin{equation}\label{eq:aggregate-normalization}
(F_1(1,1),F_2(1,1))
=\frac1\zeta\int_I\widehat f\,d\lambda
=\frac1\zeta\int_I\widehat e\,d\lambda
=(1,1).
\end{equation}
The first equality uses $B_{(1,1)}=I$.
The second follows from market clearing and the linearity
of $x\mapsto\widehat x$. The last uses
$\widehat e=1_D(1,1)$ and $\lambda(D)=\zeta$.
Let $c=\widetilde p^1-\widetilde p^2,  q=\widetilde p^1g^1-\widetilde p^2g^2$, and define
\begin{equation}\label{defn_h}
 H=\widetilde p^1(1+F_1)-\widetilde p^2(1+F_2).
\end{equation}
Since $g$ takes only the values
$(1/\widetilde p^1,0)$ and $(0,1/\widetilde p^2)$, we have
$
(\widetilde p^1g^1,\widetilde p^2g^2)\in\{(1,0),(0,1)\}.
$
Hence $q\in\{-1,1\}$. Using \eqref{eq:rectangle-aggregates} and \eqref{defn_h},
we obtain
\begin{align*}
H(v,w)
&=c+\widetilde p^1F_1(v,w)-\widetilde p^2F_2(v,w)\\
&=c+\int_0^v\int_0^w
  \bigl[\widetilde p^1g^1(t,j)-\widetilde p^2g^2(t,j)\bigr]\,dj\,dt
=c+\int_0^v\int_0^w q(t,j)\,dj\,dt.
\end{align*}At coordinates $(v,w)$, spending all income on composite good $k$
gives utility
\[
U_k(v,w)
=\{\widetilde p^k[1+F_k(v,w)]\}^{-3/2}.
\]
For almost every $i\in D$, write $(v,w)=\varphi(i)$.
If $H(v,w)>0$, then $U_2(v,w)>U_1(v,w)$, so the agent
chooses the second composite good and $q(v,w)=-1$.
If $H(v,w)<0$, then $U_1(v,w)>U_2(v,w)$, so the agent
chooses the first composite good and $q(v,w)=1$.
Individual optimality gives the sign restrictions at
$\varphi(i)$ for $\lambda_D$-almost every $i\in D$.
Since $\varphi$ is uniformly distributed on $[0,1]^2$
under $\lambda_D$, the same sign restrictions hold almost
everywhere on the square with respect to two-dimensional
Lebesgue measure. Lemma~\ref{lem:rectangle} therefore gives $|H|\le|c|$.
By \eqref{eq:aggregate-normalization},
$
H(1,1)=2\widetilde p^1-2\widetilde p^2=2c.
$
Hence $2|c|\le|c|$, so $c=0$. The lemma then gives
$q=0$ almost everywhere, contradicting $q\in\{-1,1\}$.
\revision{Thus, failure of social anonymity yields an $l$-good economy $(\mathcal E,\alpha)$ with the given agent space and social type function and no equilibrium. Consequently, statement~(ii) implies social anonymity.}
\end{proof}

\subsection{Proof of Theorem~\ref{lct-ex2}}\label{app:proof-theorem2}

The sufficiency proof uses one compact restriction of production and one approximation of the price simplex. At an auxiliary fixed point, the restriction on production does not bind,
so the firm already maximizes profit over its original production set. We initially exclude only strictly affordable improvements from household choices. Uniform tightness and compensability, together with the income condition, rule out zero limiting prices. Conditional selection then yields a limiting allocation, whose individual optimality follows from income dispersion.

\begin{proof}
(i)$\Rightarrow$(ii): \revision{Suppose that $\alpha$ satisfies social anonymity, so $\mathcal I$ is nowhere equivalent to $\mathcal G=\sigma(\alpha)$. Consider any generalized economy with the given agent space and social type function.} Write $d=l+m$, $\bar e=\int_I e\,d\lambda$, and $\Delta=\left\{p\in\mathbb R_+^d:\sum_{j=1}^d p^j=1\right\}$. 

\medskip
\noindent\emph{Step 1. Auxiliary prices, production, and household choices.}
Let
$
C:=\operatorname{rec}Y
=\{c\in\mathbb R^d:y+tc\in Y
\text{ for all }y\in Y,\ t\ge0\}.
$
Since $Y$ is closed and convex, $C$ is a closed convex cone.
Free disposal and $0\in Y$ give
$-\mathbb R_+^d\subseteq C\subseteq Y$.
Irreversibility therefore implies $C\cap(-C)=\{0\}$. Let $S=\{c\in C:\|c\|_2=1\}$. This set is nonempty and compact,
and $0\notin\operatorname{co}S$: otherwise, a convex combination
equal to zero would place a nonzero vector and its negative in $C$.
Strict separation of $0$ and the compact convex set
$\operatorname{co}S$ gives $\omega\in\mathbb R^d$ and $\eta>0$
such that $\omega\cdot s\le-\eta$ for all $s\in S$.
By scaling, $\omega\cdot c\le-\eta\|c\|_2$ for all $c\in C$.
Thus, whenever $\|\omega'-\omega\|_2<\eta$, the Cauchy--Schwarz inequality gives
$
\omega'\cdot c=\omega\cdot c+(\omega'-\omega)\cdot c \le(-\eta+\|\omega'-\omega\|_2)\|c\|_2
\le0$
 for all $c\in C$.
Hence the polar cone
$C^\circ:=\{\omega'\in\mathbb R^d:
\omega'\cdot c\le0\text{ for all }c\in C\}$
contains the open ball
$\mathcal O(\omega,\eta):=
\{\omega'\in\mathbb R^d:\|\omega'-\omega\|_2<\eta\}$.
Therefore $\operatorname{int}C^\circ\ne\varnothing$.

Choose $p^0\in\operatorname{int}C^\circ$ with $\sum_jp^{0,j}=1$. Then $p^0\gg0$ and $p^0\cdot c<0$ for every nonzero $c\in C$. For $b\in\mathbb R$, the upper-level set
$L_b:=\{y\in Y:p^0\cdot y\ge b\}$ is closed.
If it were unbounded, we could choose $y_n\in L_b$
with $\|y_n\|_2\to\infty$ and, after passing to a subsequence,
$y_n/\|y_n\|_2\to c$ with $\|c\|_2=1$.
For any $y\in Y$ and $t\ge0$, convexity gives
$(1-t/\|y_n\|_2)y+(t/\|y_n\|_2)y_n\in Y$
for all large $n$. Taking limits and using closedness yields
$y+tc\in Y$, so $c\in C$. But $p^0\cdot y_n\ge b$ implies
$p^0\cdot c\ge0$, contradicting $p^0\cdot c<0$
for nonzero $c\in C$. Hence $L_b$ is bounded and therefore compact.
Fix
$
\bar Y=\{y\in Y:p^0\cdot y\geq-p^0\cdot\bar e-1\}.
$
This set is compact and convex and contains zero. Define
\[
\bar\pi(p)=\max_{y\in\bar Y}p\cdot y,
\qquad
\Phi(p)=\argmax_{y\in\bar Y}p\cdot y,
\qquad
M=\max_{p\in\Delta}\bar\pi(p).
\]
The profit function $\bar\pi(p)$ is continuous and nonnegative, and $\Phi$ is upper hemicontinuous with nonempty, compact, and convex values.

For $0<\varepsilon<1$, let
\[
\Delta_\varepsilon=(1-\varepsilon)\Delta+\varepsilon p^0,
\qquad
\eta_\varepsilon=\varepsilon\min_jp^{0,j}>0.
\]
For $p\in\Delta$ with $p\gg0$, retain the full household budget
\[
B_i(p)=\{x\in X:p\cdot x\leq w_i(p)\},
\qquad
w_i(p)=p\cdot e(i)+\theta(i)\bar\pi(p).
\]
At $p\in\Delta_\varepsilon$, every affordable bundle satisfies
\[
\|x\|_1\leq b_\varepsilon(i)
:=\eta_\varepsilon^{-1}\bigl(\|e(i)\|_1+M\theta(i)\bigr).
\]
Let $Q_i(p,\gamma)$ consist of the bundles in $B_i(p)$ that admit no strictly affordable improvement:
\[
Q_i(p,\gamma)=
\left\{x\in B_i(p):
\text{there is no }z\in X\text{ with }
z\succ_i(\gamma)x\text{ and }p\cdot z<w_i(p)
\right\}.
\]
Since $B_i(p)$ is nonempty and compact and
$\succ_i(\gamma)$ is transitive and irreflexive with an open graph,
\citet[Theorem~3, p.~29]{hi74} guarantees a
$\succ_i(\gamma)$-maximal element of $B_i(p)$.
Every such element belongs to $Q_i(p,\gamma)$, so
$Q_i(p,\gamma)\ne\varnothing$. Budget feasibility is preserved under limits. By continuity of
income and openness of the preference graph, the inequalities
$z\succ_i(\gamma)x$ and $p\cdot z<w_i(p)$ persist locally.
Thus $Q_i$ has a closed graph. The bound $b_\varepsilon(i)$
places all values in a common compact set, so $Q_i$ is
compact-valued and upper hemicontinuous in $(p,\gamma)$.
Fix $(p,\gamma)$ and a countable dense set $X_0\subseteq X$.
Since strict preference and strict affordability are open
in the improving bundle, any such improvement can be chosen
in $X_0$. Membership in $Q_i(p,\gamma)$ is therefore determined
by the budget constraint and countably many measurable conditions.
Thus $i\mapsto Q_i(p,\gamma)$ has a measurable graph.
Its compact values and completeness of the agent space
then imply that the correspondence is measurable. Finally, every $x\in Q_i(p,\gamma)$ exhausts the budget. Otherwise, a small increase in a divisible good would be a strictly affordable improvement.

Consider the convex set
\[
\Gamma_\varepsilon
=\{\gamma\in\Gamma:0\leq\gamma^j\leq
\mathbb E(b_\varepsilon\mid\alpha)
\text{ a.e., }j=1,\ldots,d\}.
\]
The common integrable bound gives uniform integrability.
Since $\Gamma_\varepsilon$ is also weakly closed, it is
weakly compact by the Dunford--Pettis theorem.
As $\mathcal G$ is countably generated, the relative weak
topology on $\Gamma_\varepsilon$ is metrizable. The conditional-aggregation argument in Step~1 of Appendix~\ref{app:proof-theorem1} shows that
\[
\mathcal Q_\varepsilon(p,\gamma)
=\{\mathbb E(f\mid\alpha):f(i)\in Q_i(p,\gamma)\text{ a.e.}\}
\]
is upper hemicontinuous, takes values in $\Gamma_\varepsilon$, and has nonempty, convex, and weakly compact values. The price correspondence
\[
A_\varepsilon(\gamma,y)
=\argmax_{q\in\Delta_\varepsilon}
q\cdot\left(\int_I\gamma\,d\lambda-\bar e-y\right)
\]
is upper
hemicontinuous and has nonempty, compact, and convex
values in $\Delta_\varepsilon$. Fan--Glicksberg applied to
$A_\varepsilon\times\mathcal Q_\varepsilon\times\Phi$ on
$\Delta_\varepsilon\times\Gamma_\varepsilon\times\bar Y$
gives a fixed point $(p_\varepsilon,\gamma_\varepsilon,y_\varepsilon)$ and an allocation $f_\varepsilon$ such that
\[
\gamma_\varepsilon=\mathbb E(f_\varepsilon\mid\alpha),
\qquad
f_\varepsilon(i)\in Q_i(p_\varepsilon,\gamma_\varepsilon)
\quad\text{a.e.}
\]
Write $x_\varepsilon=\int_I f_\varepsilon\,d\lambda$ and
$z_\varepsilon=x_\varepsilon-\bar e-y_\varepsilon$. Integrating budget exhaustion and using $\int_I\theta\,d\lambda=1$ gives $p_\varepsilon\cdot x_\varepsilon=p_\varepsilon\cdot\bar e+\bar\pi(p_\varepsilon)=p_\varepsilon\cdot(\bar e+y_\varepsilon)$. The price correspondence therefore gives
\begin{equation}\label{eq:a2-price-inequality}
p_\varepsilon\cdot z_\varepsilon=0,
\qquad q\cdot z_\varepsilon\leq0
\quad(q\in\Delta_\varepsilon).
\end{equation}

\medskip
\noindent\emph{Step 2. The production restriction is inactive.}
Since $p^0\in\Delta_\varepsilon$ and $x_\varepsilon\geq0$,
\begin{equation}\label{eq:truncation-slack}
p^0\cdot y_\varepsilon
=p^0\cdot x_\varepsilon-p^0\cdot\bar e-p^0\cdot z_\varepsilon
\geq-p^0\cdot\bar e>-p^0\cdot\bar e-1.
\end{equation}
Suppose some $y'\in Y$ satisfies
$p_\varepsilon\cdot y'>p_\varepsilon\cdot y_\varepsilon$.
For $0<t<1$, convexity gives
$y_t:=(1-t)y_\varepsilon+ty'\in Y$.
By \eqref{eq:truncation-slack} and continuity in $t$,
$p^0\cdot y_t>-p^0\cdot\bar e-1$
for all sufficiently small $t>0$.
Thus $y_t\in\bar Y$ and
$p_\varepsilon\cdot y_t>p_\varepsilon\cdot y_\varepsilon$,
contradicting the optimality of $y_\varepsilon$ over $\bar Y$. Hence
\begin{equation}\label{eq:max_profit}
y_\varepsilon\in\argmax_{y\in Y}p_\varepsilon\cdot y,
\qquad
\bar\pi(p_\varepsilon)=\sup_{y\in Y}p_\varepsilon\cdot y.
\end{equation}

Moreover, \eqref{eq:a2-price-inequality} gives
\begin{equation}\label{eq:a2-aggregate-bound}
0\leq p^0\cdot x_\varepsilon
\leq p^0\cdot\bar e+p^0\cdot y_\varepsilon
\leq p^0\cdot\bar e+\max_{y\in\bar Y}p^0\cdot y.
\end{equation}
Since $p^0\gg0$ and $\bar Y$ is compact,
\eqref{eq:a2-aggregate-bound} implies that
$(x_\varepsilon)$ is uniformly bounded. As $\gamma_\varepsilon=\mathbb E(f_\varepsilon\mid\alpha)$ is nonnegative and has the same integral as $f_\varepsilon,$ $(\gamma_\varepsilon)$ is uniformly bounded in $L^1$. Finally, $y_\varepsilon\in\bar Y$ and $z_\varepsilon=x_\varepsilon-\bar e-y_\varepsilon$
imply that $(z_\varepsilon)$ is uniformly bounded.

Choose $\varepsilon_n\downarrow0$ and a subsequence such that
\[
p_n\to p\in\Delta,
\qquad y_n\to y\in\bar Y,
\qquad x_n\to x,
\qquad z_n\to z,
\]
where the subscript $n$ replaces $\varepsilon_n$. Applying \eqref{eq:a2-price-inequality} to
$(1-\varepsilon_n)q+\varepsilon_np^0$ and passing to the limit
gives $q\cdot z\le0$ for every $q\in\Delta$, hence $z\le0$.
By \eqref{eq:max_profit},
$p_n\cdot y_n\ge p_n\cdot y'$ for every $y'\in Y$.
Taking limits, together with $p_n\cdot z_n=0$ and
$x_n=\bar e+y_n+z_n$, yields
\begin{equation}\label{eq:a2-limit-relations}
z\le0,\qquad p\cdot z=0,\qquad
x=\bar e+y+z,\qquad
y\in\argmax_{y'\in Y}p\cdot y'.
\end{equation}
Since $y_n\in\bar Y$ and $\bar Y$ is closed, $y\in\bar Y$.
By \eqref{eq:a2-limit-relations}, $y$ maximizes profit over $Y$.
Hence $\bar\pi(p)=\pi(p)=p\cdot y<\infty$.
Moreover, \eqref{eq:max_profit} implies 
$\bar\pi(p_n)=p_n\cdot y_n\to\pi(p)$.

\medskip
\noindent\emph{Step 3. The limiting price is strictly positive.}
Let
\[
w_n(i)=p_n\cdot e(i)+\theta(i)\bar\pi(p_n),
\qquad
w(i)=p\cdot e(i)+\theta(i)\pi(p).
\]
Then $w_n(i)\to w(i)$ almost everywhere. By the resource condition, some $\widehat y\in Y$ satisfies $\bar e+\widehat y\gg0$, and hence
\begin{equation}\label{eq:a2-positive-income}
\int_I w\,d\lambda
=p\cdot\bar e+\pi(p)
\geq p\cdot(\bar e+\widehat y)>0.
\end{equation}
Let $E(p)=\{p\cdot(0^l,z):z\in\mathbb Z_+^m\}$ and define
\[
A=
\begin{cases}
\{i:w(i)>0\},
& \displaystyle\sum_{h=1}^l p^h=0,\\[2mm]
\{i:w(i)>0,\ w(i)\notin E(p)\},
& \displaystyle\sum_{h=1}^l p^h>0.
\end{cases}
\]
By \eqref{eq:a2-positive-income},
$\lambda\{i:w(i)>0\}>0$. In the second case, condition (10) and countability of $E(p)$ imply $\lambda(A)>0$ when $m\geq1$, while $E(p)=\{0\}$ when $m=0$. Thus $\lambda(A)>0$ in either case.

Suppose $p^j=0$, and write $\delta=\lambda(A)>0$. Since $f_n\ge0$ and $(x_n)$ is uniformly bounded,
Markov's inequality gives $R<\infty$ such that
$\lambda\{i:\|f_n(i)\|_1>R\}<\delta/4$ for every $n$.
Moreover, $\gamma_n=\mathbb E(f_n\mid\alpha)$ satisfies
$\int_I\|\gamma_n(i)\|_1\,d\lambda=\|x_n\|_1$.
Thus condition~(9) gives a compact set $K\subseteq\wp_+$
such that $\lambda\{i:\succ_i(\gamma_n)\notin K\}<\delta/4$
for every $n$. Thus $S_n:=\{i\in A:\|f_n(i)\|_1\le R,\,
\succ_i(\gamma_n)\in K\}$ satisfies $\lambda(S_n)\ge\delta/2$.
Hence
$\lambda(\limsup_n S_n)\ge\limsup_n\lambda(S_n)\ge\delta/2$.
Since the following properties hold outside a common null set,
we can choose $i\in\limsup_n S_n$ such that
$f_n(i)\in Q_i(p_n,\gamma_n)$ and
$p_n\cdot f_n(i)=w_n(i)$ for every $n$, and
$w_n(i)\to w(i)$.
This agent belongs to $S_n$ for infinitely many $n$.
Restricting to these indices, compactness of
$\{x\in X:\|x\|_1\le R\}$ and $K$ gives
\[
f_n(i)\to a\in X,\qquad
\succ_i(\gamma_n)\to\succ^*\in K.
\]
Budget exhaustion and income convergence give
\[
p\cdot a=\lim_n p_n\cdot f_n(i)
=\lim_n w_n(i)=w(i)>0.
\]
We construct $b\in X$ such that $b\succ^*a$ and $p\cdot b<w(i)$.
If $\sum_{h=1}^l p^h=0$, condition (8) for $\succ^*\in K$ gives $c\in\mathbb R_+^l$ such that $b=(c,0^m)\succ^*a$, and $p\cdot b=0<w(i)$.
Otherwise, $p\cdot a=w(i)\notin E(p)$ implies $p^ha^h>0$ for some divisible coordinate $h\leq l$.
Write $q^k$ for the $k$th coordinate unit vector. Strict monotonicity gives $a+q^j\succ^*a$.
For sufficiently small $t\in(0,a^h)$, openness of the strict-preference graph gives
\[
b=a+q^j-tq^h\succ^*a,\qquad
p\cdot b=w(i)-tp^h<w(i).
\]
In either case, price and income convergence imply
$p_n\cdot b<w_n(i)$ eventually along the subsequence.
Closed convergence of the complements of strict-preference
graphs also gives $b\succ_i(\gamma_n)f_n(i)$ eventually.
Otherwise, the limiting pair $(b,a)$ would belong to the
complement of $\succ^*$.
This contradicts $f_n(i)\in Q_i(p_n,\gamma_n)$.
Thus $p\gg0$. Since $z\le0$ and $p\cdot z=0$,
we obtain $z=0$ and $x=\bar e+y$.

\medskip
\noindent\emph{Step 4. Recovering the equilibrium allocation.}
All coordinates of $p_n$ eventually exceed some $\beta>0$, so budget feasibility gives the common integrable bound
\[
\|f_n(i)\|_1\leq\beta^{-1}\bigl(\|e(i)\|_1+M\theta(i)\bigr).
\]
Since $(\gamma_n)$ lies in a common weakly compact order interval,
pass to a subsequence with $\gamma_n\rightharpoonup\gamma$.
By \citet[Lemma~3]{HS2018}, there exists an integrable $f$
such that $f(i)\in\ls_n f_n(i)$ almost everywhere and
$\mathbb E(f\mid\alpha)=\gamma$.  Since $X$ is closed and $f(i)\in\ls_n f_n(i)$,
we have $f(i)\in X$ almost everywhere. Testing $\gamma_n\rightharpoonup\gamma$ against constant vectors
and using $\mathbb E(f\mid\alpha)=\gamma$ gives
\begin{equation}\label{eq:a2-market-clearing}
\int_I f\,d\lambda
=\int_I\gamma\,d\lambda
=\lim_n\int_I\gamma_n\,d\lambda
=x=\bar e+y.
\end{equation}
For almost every $i$, choose a subsequence with $f_n(i)\to f(i)$.
Taking limits in the budget inequalities gives
$p\cdot f(i)\le w(i)$.
If some $b\in X$ satisfied
$b\succ_i(\gamma)f(i)$ and $p\cdot b<w(i)$,
openness of the preference graph and convergence of prices
and incomes would imply
$b\succ_i(\gamma_n)f_n(i)$ and $p_n\cdot b<w_n(i)$
eventually along this subsequence.
This contradicts $f_n(i)\in Q_i(p_n,\gamma_n)$.
Hence $f(i)\in Q_i(p,\gamma)$ almost everywhere.

Since $p\gg0$, zero-income agents have budget $\{0\}$.
For almost every positive-income agent, $w(i)\notin E(p)$:
this follows from condition~(10) when $m\ge1$ and from
$E(p)=\{0\}$ when $m=0$.
Suppose $b\in X$ satisfies
$b\succ_i(\gamma)f(i)$ and $p\cdot b\le w(i)$.
Since $f(i)\in Q_i(p,\gamma)$,
$p\cdot b=w(i)\notin E(p)$.
Thus $b^h>0$ for some divisible coordinate $h\le l$.
Writing $q^h$ for the $h$th coordinate unit vector,
openness of the preference graph gives, for sufficiently
small $t\in(0,b^h)$,
\[
b-tq^h\succ_i(\gamma)f(i),
\qquad
p\cdot(b-tq^h)=w(i)-tp^h<w(i),
\]
contradicting $f(i)\in Q_i(p,\gamma)$.
Hence $f$ is individually optimal almost everywhere.
Together with profit maximization in
\eqref{eq:a2-limit-relations} and market clearing in
\eqref{eq:a2-market-clearing}, this proves that
$(f,p,y)$ is an equilibrium.

\medskip
(ii)$\Rightarrow$(i): \revision{The exchange economies of Theorem~\ref{lct-ex1} with the given social type function $\alpha$ can be embedded by setting $m=0$ and $Y=-\mathbb R_+^l$, without changing $\alpha$.}\footnote{Define $x\succ_i(\gamma)z$ iff $u_i(x,\gamma)>u_i(z,\gamma)$, and take $\theta\equiv1$. The assumptions on $u_i$ imply the required properties of the induced preferences $\succ_i$. The income-dispersion condition is vacuous when $m=0$.}
Condition~(8) then follows from strict monotonicity.
In any generalized equilibrium, strict monotonicity gives
$p\gg0$, so profit maximization implies $y=0$.
Such an equilibrium is therefore an exchange equilibrium.
The counterexample in the necessity proof of
Theorem~\ref{lct-ex1} then proves the claim.
\end{proof}

\clearpage
\setcounter{page}{1}
\setcounter{section}{2}
\setcounter{subsection}{0}
\phantomsection
\addcontentsline{toc}{section}{Supplemental Appendix}
\begin{center}
{\large\bfseries
Supplement to ``Large Economies with Social Types''\par}
\medskip
{\large
Wei He, Yeneng Sun, Bin Wu, Nicholas C. Yannelis\par}
\end{center}

%
\subsection{Proof of Claim~\ref{claim:car-equilibrium}}
\label{app:car-proof}

We first verify the assumptions of Section~\ref{glc}, and then check the proposed equilibrium.

\medskip
\noindent\textit{Step~1. The population, production, and income.}
The agent space is the completed Lebesgue unit square,
and $\mathcal G$ is generated by $(v,w)\mapsto v$.
By \citet[Lemma~1, p.~780]{HSS2017}, the Borel product
$\sigma$-algebra is nowhere equivalent to $\mathcal G$.
Since completion preserves this property,
$\mathcal I$ is nowhere equivalent to $\mathcal G$.

The endowment and ownership mappings are measurable and integrable, with the aggregates stated in the example. The production set $Y$ is closed, convex, contains zero, and permits free disposal. If both $y$ and $-y$ belong to $Y$, the inequalities on their first coordinates give $y^1=0$, and those on their coordinate sums then give $y^2=0$. Hence $Y\cap(-Y)=\{0\}$. Moreover,
\[
\int_I e\,d\lambda+\left(-\frac12,\frac12\right)
=\left(2,\frac12\right)\gg0,
\]
which verifies the aggregate resource condition.

For every $p=(p^1,p^2)\geq0$, maximal profit is
\[
\pi(p)=\sup_{y\in Y}p\cdot y
=\frac12(p^2-p^1)^+,
\qquad a^+=\max\{a,0\}.
\]
Indeed, a plan with $y^2<0$ yields nonpositive profit. If $y^2=q\geq0$, then $q\leq\frac12$ and $y^1\leq-q$, so $p\cdot y\le (p^2-p^1)q$. For every $q\in[0,1/2]$, the plan $(-q,q)$ belongs to $Y$
and earns $(p^2-p^1)q$. Thus $q=1/2$ attains the maximum
if $p^2>p^1$, and $q=0$ does so if $p^2\le p^1$.

The income of household $(v,w)$ is
\[
p\cdot e(v,w)+\theta(v,w)\pi(p)
=p^1(2+v)
+\left(v+\frac12\right)\frac12(p^2-p^1)^+.
\]
As a function of $v$, this is affine with slope
\[
p^1+\frac12(p^2-p^1)^+>0
\qquad\text{whenever }p\geq0\text{ and }p\neq0.
\]
Since income is strictly increasing in $v$, each income level
corresponds to at most one wealth rank. As $v$ is uniformly
distributed, the income distribution assigns zero probability
to every singleton, as required by condition~(10).

\medskip
\noindent\textit{Step~2. Preferences.} 
For fixed $v>0$, the function
$v^{-1}\mathbf 1_{[0,v]\times[0,1]}$ belongs to
$L_\infty(I,\mathcal G,\lambda)$. Hence
\[
h\to \frac1v\int_{[0,v]\times[0,1]}h^2\,d\lambda
\]
is a bounded linear functional on
$L_1((I,\mathcal G,\lambda);\mathbb R^2)$,
and its restriction $r_v$ to $\Gamma$ is weakly continuous. At $v=0$, $r_0\equiv 1/2$.
Hence $\eta_v=\min\{1,r_v\}$ is weakly continuous for each $v$,
and the utility formula gives joint continuity of $u_{(v,w)}$ in $(x,\gamma)$.
For fixed $\gamma$, the cumulative integral defining $r_v(\gamma)$
is continuous in $v$, so $v\mapsto\eta_v(\gamma)$ is continuous
on $(0,1]$. With its specified value at zero, this map is
measurable on $[0,1]$. The utility is consequently jointly measurable in the household and its consumption bundle. In particular, its strict-preference graph has the required measurability. The utility representation gives irreflexivity and transitivity, and joint continuity gives an open strict-preference graph in consumption and the externality. Strict monotonicity follows from
\[
\frac12\leq\frac32-\eta_v(\gamma)\leq\frac32
\]
and the strict increase of $q\mapsto\log(1+q)$.

For uniform tightness, define
\[
U_\eta(x)
=x^1+\left(\frac32-\eta\right)
\frac{\log(1+x^2)}{\log 2},
\qquad 0\leq\eta\leq1,
\]
and let $\succ_\eta$ be its strict preference relation. We show that $\eta\mapsto\succ_\eta$ is continuous in the topology of closed convergence. Write
\[
F_\eta=\{(x,z)\in X\times X:U_\eta(x)\leq U_\eta(z)\}
\]
for the complement of the strict-preference graph. Let $\eta_n\to\eta$. If $(x_k,z_k)\in F_{\eta_{n_k}}$ and
$(x_k,z_k)\to(x,z)$, joint continuity gives
$U_\eta(x)\le U_\eta(z)$, so $(x,z)\in F_\eta$.
Thus $\limsup_n F_{\eta_n}\subseteq F_\eta$.
Conversely, fix $(x,z)\in F_\eta$ and set
\[
\delta_n=\max\{0,U_{\eta_n}(x)-U_{\eta_n}(z)\},
\qquad
z_n=(z^1+\delta_n,z^2).
\]
Since $U_\eta(x)\le U_\eta(z)$, continuity implies
$\delta_n\to0$. Only the divisible coordinate is increased,
so $z_n\in X$, and the utility formula gives
$
U_{\eta_n}(z_n)
=U_{\eta_n}(z)+\delta_n
\ge U_{\eta_n}(x).
$
Hence $(x,z_n)\in F_{\eta_n}$ and $(x,z_n)\to(x,z)$,
showing $F_\eta\subseteq\liminf_n F_{\eta_n}$.
Thus $\liminf_n F_{\eta_n}=\limsup_n F_{\eta_n}=F_\eta$.
By condition~(7) of Section~\ref{glc}, this is precisely
$\succ_{\eta_n}\to\succ_\eta$ in $\wp$.
Hence $\eta\mapsto\succ_\eta$ is continuous, and its image
$K=\{\succ_\eta:\eta\in[0,1]\}$ is compact.
For any $x\in X$, choosing
\[
c>x^1+\frac{3\log(1+x^2)}{2\log2}
\]
gives $U_\eta(c,0)>U_\eta(x)$ for every $\eta\in[0,1]$.
Thus every preference in $K$ satisfies condition~(8).
Together with strict monotonicity, this gives $K\subseteq\wp_+$.
Since $v\mapsto\eta_v(\gamma)$ is measurable and
$\eta\mapsto\succ_\eta$ is continuous, the conditional
preference assignment is measurable.
All conditional preferences lie in the same compact set $K$,
so $\mu_\gamma(K)=1$ for every $\gamma\in\Gamma$.
This verifies uniform tightness in condition~(9).

\medskip
\noindent\textit{Step~3. Equilibrium.}
Let $A=[0,1]\times[0,\frac12]$ and $\gamma^*=\mathbb E(f^*\mid\alpha)$. Since $\mathbb E(\mathbf1_A\mid\alpha)=\frac12$, the proposed allocation satisfies
$\gamma^*(v,w)=\left(\frac32+v,\frac12\right)$. Hence $r_v(\gamma^*)=\eta_v(\gamma^*)=\frac12$ for every $v\in[0,1]$.

At $p^*=(1,1)$, maximal profit is zero and income is
$b=2+v$. For every affordable bundle $(x^1,q)\in X$,
\[
u_{(v,w)}((x^1,q),\gamma^*)
=x^1+\frac{\log(1+q)}{\log 2}
\le b-q+\frac{\log(1+q)}{\log 2}
\le b,
\]
where the last inequality follows from $1+q\le 2^q$
for $q\in\mathbb Z_+$.
Both assigned bundles, $(b,0)$ and $(b-1,1)$, attain
this bound and are therefore optimal.

Finally,
\[
\int_I f^*\,d\lambda
=\left(2,\frac12\right)
=\int_I e\,d\lambda+y^*.
\]
The plan $y^*=(-\frac12,\frac12)$ belongs to $Y$ and satisfies $p^*\cdot y^*=0=\pi(p^*)$. Thus, individual optimality, profit maximization, and market clearing all hold, proving the claim.

\subsection{Proof of the claim in Remark~\ref{rmk-tightness-essential}
\label{app:tightness-proof}}
\noindent\textit{Step 1. Construction and regularity.} Let $I=[0,1]^2$, with completed Lebesgue measure, and write an agent as $(v,w)$. There are two divisible goods, $e(v,w)=(1,1)$, and $\alpha(v,w)=v\in[0,1]$. Thus $\mathcal G=\sigma(\alpha)$ records the first coordinate. For $\gamma\in\Gamma$, define
\[
A(\gamma)=\int_I\gamma^1\,d\lambda,
\qquad B(\gamma)=\int_I v\gamma^1(v,w)\,d\lambda,
\qquad
c(\gamma)=
\begin{cases}
\dfrac{(A(\gamma)-1/2)_+}{B(\gamma)^2},&A(\gamma)>0,\\[6pt]
0,&A(\gamma)=0,
\end{cases}
\]
where $a_+=\max\{a,0\}$. Utilities are $u_{(v,w)}(x,\gamma)=x^1+[1+vc(\gamma)]x^2$.
The externality increases the relative value of the second good, with a larger response at higher social types.

Both $A$ and $B$ are weakly continuous. Since $\gamma^1\ge0$
and $v>0$ almost everywhere, $A(\gamma_0)>0$ implies $B(\gamma_0)>0$.
On the weak neighborhood $\{B>B(\gamma_0)/2\}$ of $\gamma_0$,
the denominator is bounded away from zero, so
$c=(A-1/2)_+/B^2$ is continuous. This includes $A(\gamma_0)=1/2$.
If $A(\gamma_0)=0$, then $c$ vanishes on the weak neighborhood
$\{A<1/2\}$ of $\gamma_0$. Utility is therefore jointly continuous, measurable in the agent, and strictly monotone. Preferences are linear and hence convex. \revision{The agent space and social type function are the same as in Section~\ref{subsec:general-example}. Social anonymity therefore holds, as shown in Step~1 of the proof of Claim~\ref{claim:car-equilibrium}.}

\medskip
\noindent\textit{Step 2. Nonexistence of equilibrium.}
Suppose that $(f,p)$ is an equilibrium and write $\gamma=\mathbb E(f\mid\alpha)$. Strict monotonicity rules out a zero price for either good. Normalize $p=(s,1-s)$ with $0<s<1$, so every household has income one. Market clearing gives $A(\gamma)=1$ and hence $c=c(\gamma)>0$. The two budget vertices yield utilities $1/s$ and
$(1+vc)/(1-s)$. Thus an agent uniquely chooses good~1
when
\[
v<\tau:=\frac{1-2s}{sc},
\]
and good~2 when $v>\tau$. Since $v$ is uniformly distributed, the indifferent households
$\{(v,w)\in I:v=\tau\}$ have measure zero, so their choices do
not affect aggregate demand. If $\tau\le0$, aggregate demand
for good~1 is zero. If $\tau\ge1$, aggregate demand for good~2
is zero. Since each good has unit supply, market clearing
requires $0<\tau<1$. Clearing the market for good~1 then gives
\[
1=\int_I f^1\,d\lambda
 =\frac1s\lambda\{(v,w)\in I:v<\tau\}
 =\frac{\tau}{s}.
\]
Hence $\tau=s$. Since $v$ is bounded and
$\mathcal G$-measurable,
\[
B(\gamma)
=\int_I v\,\mathbb E(f^1\mid\alpha)\,d\lambda
=\int_I vf^1\,d\lambda
=\frac1s\int_0^s v\,dv
=\frac{s}{2}.
\]
It follows that
\[
c=\frac{1/2}{(s/2)^2}=\frac{2}{s^2},
\qquad
cs^2=1-2s,
\]
where the last equality follows from $\tau=s$.
Thus $2=1-2s$, a contradiction.

\medskip
\noindent\textit{Step 3. Failure of tightness and the ordinal condition.}
Consider
\[
\gamma_n(v,w)=\bigl(n\mathbf1_{\{v<1/n\}},1\bigr),
\qquad n\ge1.
\]
Each $\gamma_n$ is $\mathcal G$-measurable and satisfies
$\int_I\gamma_n\,d\lambda=(1,1)$, so it is itself a
feasible allocation inducing the externality $\gamma_n$.
Moreover, $\|\gamma_n\|_{L_1} =2$, $A(\gamma_n)=1$, $B(\gamma_n)=1/(2n)$, and $c(\gamma_n)=2n^2$. Let $K\subset\wp_+$ be a nonempty compact set in the
utility space of Section~\ref{subsec-setup}. Evaluation at $(0,1)$ is continuous in the compact-open
topology. Since $K$ is nonempty and compact, this map
attains a finite maximum:
\[
b_K:=\max_{u\in K}u(0,1)<\infty.
\]
Membership of $u_{(v,w)}(\cdot,\gamma_n)$ in $K$ therefore
requires $1+2vn^2\le b_K$. Consequently,
\[
\mu_{\gamma_n}(K)
\le\lambda\{(v,w):1+2vn^2\le b_K\}
\le\frac{(b_K-1)^+}{2n^2}
\to0.
\]
Thus uniform tightness fails. 

To test Inequality \eqref{eq:ordinal-stability} in Remark~\ref{rmk-cover}, take $M=2$, $R=1$,
$\epsilon=1/2$, $x=(0,1)$, and $q^1=(1,0)$.
For any fixed $\eta\in(0,1)$,
\[
u_{(v,w)}((1-\eta)x+q^1,\gamma_n)
-u_{(v,w)}(x,\gamma_n)
=1-\eta(1+2vn^2).
\]
The set of agents for whom this difference is positive
has measure
\[
\min\left\{1,\frac{1-\eta}{2\eta n^2}\right\}
\to 0.
\]
For every $\eta\in(0,1)$, choose $n$ with $n^2>(1-\eta)/\eta$.
Then fewer than half of the households satisfy the comparison even
for $x=(0,1)$ and $q^1=(1,0)$. Hence no $\eta$ works for every
$\gamma_n$, as Inequality \eqref{eq:ordinal-stability} in Remark~\ref{rmk-cover}  requires.
Although $c$ is continuous, the statistics
$(A(\gamma_n),B(\gamma_n))$ approach the unattainable
point $(1,0)$, and $c(\gamma_n)\to\infty$.

\subsection{Proof of the claim in Remark~\ref{rmk-transformed}}
\label{app:proof-transformed}

\revision{Suppose that $\alpha$ satisfies social anonymity, so $\mathcal I$ is nowhere equivalent to $\mathcal G=\sigma(\alpha)$.}
We use the notation and approximations from the sufficiency proof
of Theorem~\ref{lct-ex2} in Appendix~\ref{app:proof-theorem2}.
Write $\zeta$ for the transformed externality, and let
$Q_i(p,\zeta)$ be the household correspondence defined there,
with preferences evaluated at $\zeta$. We keep track of consumption
and transformed consumption jointly.

For each $0 < \varepsilon < 1$, define
\[
Z_\varepsilon
=
\left\{
\zeta\in L_1((I,\mathcal G,\lambda);\mathbb R^k):
\|\zeta(i)\|
\le a+b\,\mathbb E(b_\varepsilon\mid\alpha)(i)
\ \text{a.e.}
\right\}.
\]
By the argument for $\Gamma_\varepsilon$ in Step~1 of
Appendix~\ref{app:proof-theorem1}, $Z_\varepsilon$ is convex and
weakly compact, and its relative weak topology is metrizable.
For $p\in\Delta_\varepsilon$, every $x\in Q_i(p,\zeta)$ satisfies
$\|x\|_1\le b_\varepsilon(i)$ and hence
$\|\tau(x)\|\le a+b\,b_\varepsilon(i)$. For every measurable selection $f$ of $Q_i(p,\zeta)$,
conditional Jensen's inequality gives
\[
\bigl\|\mathbb E(\tau\circ f\mid\alpha)\bigr\|
\le
\mathbb E(\|\tau\circ f\|\mid\alpha)
\le a+b\,\mathbb E(b_\varepsilon\mid\alpha).
\]
Hence $\mathbb E(\tau\circ f\mid\alpha)\in Z_\varepsilon$. By Step~1 of Appendix~\ref{app:proof-theorem2}
and continuity of $x\mapsto(x,\tau(x))$, the correspondence
\[
H_i(p,\zeta)=\{(x,\tau(x)):x\in Q_i(p,\zeta)\}
\]
has nonempty compact values, is measurable in $i$ for each $(p,\zeta)$,
and is upper hemicontinuous in $(p,\zeta)$ for almost every $i$.
The budget bound $\|x\|_1\le b_\varepsilon(i)$ from the same step gives
$\|x\|_1+\|\tau(x)\|\le a+(1+b)b_\varepsilon(i)$, so $H_i$ is integrably bounded. 
Since $\mathcal I$ is nowhere equivalent to $\mathcal G$ and
$\Delta_\varepsilon\times Z_\varepsilon$ is metrizable,
we may apply
\citet[Theorems~2--4]{HS2018} to the joint correspondence $H_i$. Its conditional
expectation correspondence
\[
\widehat Q_\varepsilon(p,\zeta)
=\bigl\{(\mathbb E(f\mid\alpha),
          \mathbb E(\tau\circ f\mid\alpha)):
       f \text{ is a measurable selection of }Q_i(p,\zeta)\bigr\}
\]
is upper hemicontinuous and has nonempty, convex, weakly compact
values in $\Gamma_\varepsilon\times Z_\varepsilon$.

Let $D_\varepsilon=\Delta_\varepsilon\times\Gamma_\varepsilon
\times Z_\varepsilon\times\bar Y$. By Step~1 of Appendix~~\ref{app:proof-theorem2}
and the preceding argument, $D_\varepsilon$ is nonempty, compact,
and convex in the product of the Euclidean and weak $L_1$ topologies.
Define $\Psi_\varepsilon:D_\varepsilon\rightrightarrows D_\varepsilon$ by
\[
\Psi_\varepsilon(p,\gamma,\zeta,y)
=A_\varepsilon(\gamma,y)\times
 \widehat Q_\varepsilon(p,\zeta)\times\Phi(p).
\]
Here $A_\varepsilon:\Gamma_\varepsilon\times\bar Y
\rightrightarrows\Delta_\varepsilon$ and
$\Phi:\Delta_\varepsilon\rightrightarrows\bar Y$ have the properties
established in Step~1 of Appendix~\ref{app:proof-theorem2}. Thus all three factors are
upper hemicontinuous with nonempty compact convex values.
Their finite product has the same properties, so Fan--Glicksberg
gives a fixed point. The resulting fixed point is represented
by an allocation $f_\varepsilon$ satisfying
\[
\gamma_\varepsilon=\mathbb E(f_\varepsilon\mid\alpha),
\qquad
\zeta_\varepsilon=\mathbb E(\tau\circ f_\varepsilon\mid\alpha),
\qquad
f_\varepsilon(i)\in Q_i(p_\varepsilon,\zeta_\varepsilon)
\quad\text{a.e.}
\]
Budget exhaustion and the price fixed-point condition give the
production and aggregate-consumption conclusions of Step~2 in Appendix~\ref{app:proof-theorem2}.
In particular, production maximizes profit over $Y$ and
$\sup_\varepsilon\|f_\varepsilon\|_{L_1}<\infty$. Moreover,
\[
\|\zeta_\varepsilon\|_{L_1}
\le \int_I\|\tau(f_\varepsilon(i))\|\,d\lambda(i)
\le a+b\|f_\epsilon\|_{L_1}.
\]
Choose $\varepsilon_n\downarrow0$. As in Step~2 of
Appendix~\ref{app:proof-theorem2}, pass to a subsequence such that
$p_n\to p$, $y_n\to y$, and $x_n:=\int_I f_n\,d\lambda\to x$.
Set $z=x-\bar e-y$. Since $(\zeta_n)$ is norm bounded,
condition~(9) applies, and the boundary-price argument in
Step~3 gives $p\gg0$. By \eqref{eq:a2-limit-relations},
$y$ maximizes profit at $p$, $z\le0$, and $p\cdot z=0$.
Thus $z=0$ and $x=\bar e+y$.

Since $p_n\to p\gg0$, the budget bound in Step~4 of Appendix~\ref{app:proof-theorem2}
gives an integrable $g$ such that $\|f_n(i)\|_1\le g(i)$
after discarding finitely many terms. Let $F_n=(f_n,\tau\circ f_n)$.
Then $\|F_n(i)\|\le a+(1+b)g(i)$ in the product norm. Conditional
Jensen's inequality gives a common integrable bound for
$\mathbb E(F_n\mid\alpha)$, so along a subsequence
\[
(\gamma_n,\zeta_n)=\mathbb E(F_n\mid\alpha)
\rightharpoonup(\gamma,\zeta).
\]
By nowhere equivalence and \citet[Lemma~3]{HS2018}, there is an
integrable pair $(f,h)$ such that
\[
(f(i),h(i))\in\operatorname{Ls}_n F_n(i)\quad\text{a.e.},
\qquad
\mathbb E((f,h)\mid\alpha)=(\gamma,\zeta).
\]

For almost every $i$, choose a subsequence $n_j=n_j(i)$ such that
\[
(f_{n_j}(i),\tau(f_{n_j}(i)))\to(f(i),h(i)).
\]
Since $X$ is closed and $\tau$ is continuous, $f(i)\in X$ and
$h(i)=\tau(f(i))$. Hence $\zeta=\mathbb E(\tau\circ f\mid\alpha)$.
Testing $\gamma_n\rightharpoonup\gamma$ against constant vectors gives
\[
\int_I f\,d\lambda
=\int_I\gamma\,d\lambda
=\lim_n\int_I f_n\,d\lambda
=\bar e+y.
\]
Along the same individual subsequence, $p_{n_j}\to p$,
$\zeta_{n_j}\rightharpoonup\zeta$, and $f_{n_j}(i)\to f(i)$.
Since $f_{n_j}(i)\in Q_i(p_{n_j},\zeta_{n_j})$, the closed graph
of $Q_i$ gives $f(i)\in Q_i(p,\zeta)$. The budget-boundary argument
in the last paragraph of Step~4 of Appendix~\ref{app:proof-theorem2} now gives individual
optimality almost everywhere. Together with profit maximization
and market clearing, this proves the claim.
\end{document}